\documentclass[aps,prl,reprint,superscriptaddress,longbibliography]{revtex4-2}
\usepackage{amsmath,amssymb,bm,graphicx,hyperref}
\hypersetup{colorlinks=true,citecolor=blue,urlcolor=blue,linkcolor=blue}
\newcommand{\ii}{\mathrm{i}}
\newcommand{\Dret}{\mathsf D_{\rm ret}^{+}}
\newcommand{\Op}{\operatorname{Op}^{W}}

\begin{document}

\title{Collisionless Resonances Set the Size of a Weyl Exceptional Ring}

\author{Xianhao Rao}
\email{rrxxhh@mail.ustc.edu.cn}
\affiliation{Xeonova Ltd., Hefei 230093, People's Republic of China}

\author{Hong Li}
\email{honglee@ustc.edu.cn}
\affiliation{Xeonova Ltd., Hefei 230093, People's Republic of China}
\affiliation{School of Nuclear Science and Technology, University of Science and Technology of China, No. 443 Huangshan Road, Hefei, Anhui, People's Republic of China}

\author{Xuan Sun}
\email{xsun@ustc.edu.cn}
\affiliation{Xeonova Ltd., Hefei 230093, People's Republic of China}
\affiliation{School of Nuclear Science and Technology, University of Science and Technology of China, No. 443 Huangshan Road, Hefei, Anhui, People's Republic of China}

\date{\today}

\begin{abstract}
Eliminating continuum degrees of freedom converts conservative dynamics into a dispersive, non-Hermitian response whose topology need not represent the causally selected kinetic dynamics.  In collisionless magnetized plasma, retarded Landau and cyclotron resonances damp two Weyl modes differently and replace a cold Weyl point by an exceptional ring.  Axial roots and root-normalized polarization mixing predict its off-axis radius, $R_{\rm EP}^{\rm pred}=|\Delta\Gamma|/(2|v_{\rm mix}|)$, before any double root is computed.  Across five nonrelativistic scans spanning a 310-fold range of differential damping, $R_{\rm EP}^{\rm pred}$ agrees with the nonlinear double-root radius $R_{\rm EP}^{\rm full}$ within $0.12\%$.  The selected retarded kinetic-response bundle is a continuous graph over the electric root line bundle and has the same first Chern class under stated conditions.  The cold interface mode continues to a localized Keldysh pole with one oriented crossing in the tested line-gap window, consistent with the inherited charge.
\end{abstract}

\maketitle

Topological waves are commonly classified by closed eigenvalue problems.  Continuous media pose a different problem because attenuation can emerge only after microscopic degrees of freedom are eliminated.  The resulting response is dispersive and non-Hermitian, and its complex roots represent resonances of an underlying conservative generator.  It is then unclear whether topology of the reduced response still characterizes the causally selected kinetic response.  The same structural issue can arise when a conservative continuum is integrated out to generate a causal self-energy, as in dispersive electromagnetic media~\cite{RamanFan2010,Silveirinha2015}.

Phenomenological loss can expand a Dirac or Weyl point into an exceptional ring.  The parent Chern charge may remain on a line-gapped enclosing surface~\cite{Zhen2015,Xu2017,Cerjan2019,McGuinness2020,Bergholtz2021}.  Line and point gaps support distinct non-Hermitian classifications~\cite{Kawabata2019}, while frequency-dependent responses require nonlinear spectral topology~\cite{KotzTimm2023,Isobe2024,Yoshida2025}.  Charge--spectral-flow relations are also known for Hermitian phase-space symbols~\cite{Delplace2017,Faure2023,QinFu2023} and certain frequency-linear non-Hermitian deformations~\cite{Jezequel2023}.  These results do not determine what happens when the imaginary self-energy is generated by a continuum rather than imposed as a linewidth.

Kinetic theory makes that distinction concrete.  Eliminating velocity space produces a causal, frequency-dependent self-energy, while the unreduced Vlasov equation retains free-streaming homogeneous solutions.  Its complex poles arise through retarded continuation of conservative Maxwell--Vlasov dynamics~\cite{Morrison1980,VanKampen1955,Case1959,Ramos2019}.  Subsystem reductions have a similar self-energy structure~\cite{Huang2026}, and linear gyrokinetic spectra can contain exceptional points~\cite{Kammerer2008}.  Two questions remain: can wave--particle resonances quantitatively determine exceptional geometry, and does the reduced electric topology represent the selected kinetic response?

Magnetized plasma supplies a controlled test because its cold Hermitian limit contains a known Weyl crossing between longitudinal and right-hand circularly polarized waves~\cite{Gao2016,FuQin2021}.  The associated interface mode and phase-space charge have been characterized in cold-plasma theory~\cite{FuQin2022,QinFu2023,Faure2023}, with complementary charge diagnostics available for continuum formulations~\cite{Fonseca2024}.  Warm-fluid, dissipative, screened-continuum, and parity-time-symmetric plasma models produce related topological or exceptional waves~\cite{Rao2025,Billings2026,FuQin2024,Rao2026,Wang2020,Shastri2020,Yan2021}, but they do not resolve both kinetic questions above.

Landau and cyclotron resonances generate a differential damping $\Delta\Gamma$, which sets the radius of a Weyl exceptional ring (WER).  The axial roots and their root-normalized polarization-mixing matrix are determined before any off-axis exceptional root is computed.  Retarded selection then defines a kinetic-response graph over the electric root line bundle.  The inherited charge is finally compared with a finite-window oriented crossing count of a localized Keldysh pole.  This endpoint comparison is numerical and does not establish index invariance throughout the temperature--momentum plane.

\emph{Exact equilibrium and causal response.---}
Take a uniform field $\bm B_0=B_0\hat{\bm z}$ and let $\Omega_s=q_sB_0/m_s$ be the signed cyclotron frequency of species $s$.  Two Maxwellian components, labeled by $a=1,2$, give the stationary distribution
\begin{equation}
\begin{aligned}
 f_{0s}(x,\bm v)&=\sum_{a=1}^{2}N_{sa}\!\left(x+\frac{v_y}{\Omega_s}\right)
 \frac{e^{-v^2/v_{tsa}^{2}}}{\pi^{3/2}v_{tsa}^{3}},
 &v_{tsa}^{2}=\frac{2T_{sa}}{m_s}.
\end{aligned}
\label{eq:eqm}
\end{equation}
Both arguments of $f_{0s}$ are single-particle invariants, so Eq.~(\ref{eq:eqm}) is an exact stationary Vlasov equilibrium.  The profiles produce a monotone error-function density from $n_-=6.50\times10^{18}\,{\rm m}^{-3}$ to $n_+=1.70\times10^{18}\,{\rm m}^{-3}$ over $L=0.10\,{\rm m}$.  Two Maxwellian components with positive density weights, $T_{e1}=0.8\,{\rm keV}$ and $T_{e2}=8\,{\rm keV}$, keep the prescribed thermal-pressure moment constant.  Electron--ion charge and current cancellation permit $\bm E_0=0$ and uniform $\bm B_0$.  This exact equilibrium separates resonant attenuation from background evolution; its construction and ion completion are given in the Supplemental Material (SM)~\cite{SupplementalMaterial}.

At fixed $X=x/L$, define $w=z/\omega_{ce}$ and $\bm u=c\bm k/\omega_{ce}$, where $z$ is complex frequency and $\omega_{ce}=eB_0/m_e>0$.  Eliminating velocity space with retarded boundary conditions gives
\begin{equation}
\begin{aligned}
 \Dret(w,\bm u,X)&=\bm\varepsilon_{\rm ret}^{+}(w,\bm u,X)\\
 &\quad-w^{-2}\bigl(u^2\bm 1-\bm u\bm u^{\mathsf T}\bigr),\\
 \det\Dret(w,\bm u,X)&=0.
\end{aligned}
\label{eq:dispersion}
\end{equation}
The tensor $\bm\varepsilon_{\rm ret}^{+}$ follows from the magnetized Vlasov susceptibility~\cite{Stix1992,Verscharen2018}.  Its parallel integral uses the retarded plasma dispersion function continued into $\operatorname{Im}z<0$~\cite{FriedConte1961}.  No collision frequency, Pad\'e closure, or resonance broadening is introduced.  The collisionless initial-value problem remains conservative; non-Hermiticity appears in its analytically continued retarded response and resonance poles.  Because $\bm\varepsilon_{\rm ret}^{+}$ depends on $w$, Eq.~(\ref{eq:dispersion}) is a nonlinear root problem.  We retain its frequency derivative for multiplicity, residue, and Keldysh-pole tests~\cite{Beyn2012}; formulas and convergence checks are in the SM~\cite{SupplementalMaterial}.

\emph{Differential damping sets the ring.---}
In the cold limit at $u_z=1$, the Langmuir and right-hand circularly polarized ($R$) branches meet at $w_c=(\sqrt5-1)/2$, $X_c\simeq0.144$, and $u_\perp=(u_x^2+u_y^2)^{1/2}=0$~\cite{FuQin2021,QinFu2023}.  We use fields proportional to $\exp(\ii\bm k\cdot\bm x-\ii zt)$; for $\bm B_0=B_0\hat{\bm z}$, the $R$ branch has $\bm E\propto(\hat{\bm x}+\ii\hat{\bm y})/\sqrt2$.  On the symmetry axis the kinetic matrix separates into scalar longitudinal and $R$-polarized equations, $m_\parallel(w,X)=0$ and $m_+(w,X)=0$.  Thus $u_\perp=0$ identifies the unmixed parent crossing, while transverse momentum supplies the off-axis polarization mixing.

The two polarizations weight the resonances $z-k_zv_z-\ell\Omega_s=0$ differently.  At $u_\perp=0$, the $k_\perp\to0$ Bessel-function limits retain the $\ell=0$ harmonic in the longitudinal sector and the fundamental cyclotron harmonic in the $R$ sector.  With the signed convention $\Omega_e=-\omega_{ce}$, the $R$-polarized branch couples resonantly to the $\ell=-1$ electron-cyclotron harmonic, whereas the longitudinal branch couples to the $\ell=0$ Landau resonance.

For weak damping, write $w_j=w_{r,j}-\ii\Gamma_j$, where $\Gamma_j=-\operatorname{Im}w_j$ is dimensionless and the physical damping rate is $\omega_{ce}\Gamma_j$.  The contribution of component $a$ to branch $j$ is
\begin{equation}
 \Gamma_j^{(a)}=
 \frac{\operatorname{Im}\chi_j^{(a)}(w_{r,j})}
 {\partial_w\operatorname{Re}m_j(w_{r,j})},
 \qquad j\in\{+,\parallel\}.
\label{eq:resonant-damping}
\end{equation}
Equation~(\ref{eq:resonant-damping}) relates the imaginary root displacement to the resonant particles and the dispersive wave energy stored in the root derivative.  Direct Landau-contour evaluation agrees with the complex roots without a regulator.

\begin{figure*}[t]
 \includegraphics[width=\textwidth]{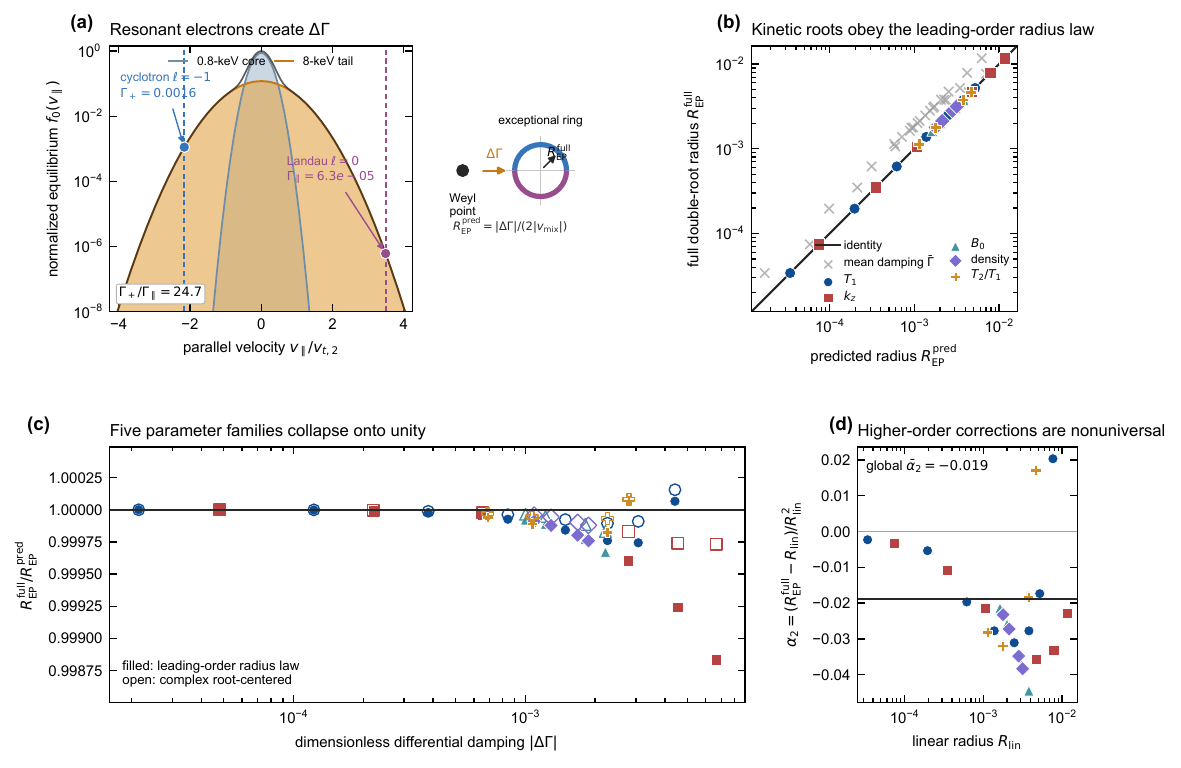}
 \caption{\label{fig:bulk}Microscopic differential damping determines the geometry of the WER.  (a) The $R$-branch $\ell=-1$ electron-cyclotron resonance and the longitudinal $\ell=0$ Landau resonance sample different parts of the two-Maxwellian equilibrium and produce unequal axial damping rates.  The displayed $\Gamma=-\operatorname{Im}w$ are dimensionless.  (b) The axial prediction $R_{\rm EP}^{\rm pred}$ is compared with the off-axis double-root radius $R_{\rm EP}^{\rm full}$ at 26 parameter points, without fitting the ring.  Gray crosses replace differential damping by mean damping.  (c) The ratio $R_{\rm EP}^{\rm full}/R_{\rm EP}^{\rm pred}$ across variations of temperature, $k_z$, $B_0$, density, and temperature ratio.  Filled symbols use Eq.~(\ref{eq:radius-law}); open symbols include the complex root-centered correction.  (d) The parameter-dependent residual delimits the leading-order regime.}
\end{figure*}

At the position where the two real frequencies coincide, the baseline roots are $w_+=0.617784-1.551\times10^{-3}\ii$ and $w_\parallel=0.617784-6.29\times10^{-5}\ii$.  Their differential damping is $\Delta\Gamma=1.488\times10^{-3}$ and their mean is $\bar\Gamma=8.07\times10^{-4}$.  The 8-keV component supplies more than $99.9\%$ of the $R$-branch damping [Fig.~\ref{fig:bulk}(a)].  The mean sets the common decay, while $\Delta\Gamma$ opens the ring.

The resonant velocities explain this selectivity.  In units of the 8-keV thermal speed, the $R$-branch cyclotron resonance lies at $v_\parallel/v_{t,2}\simeq-2.16$, where the distribution retains appreciable weight.  The longitudinal resonance lies at $v_\parallel/v_{t,2}\simeq3.49$.  For the 0.8-keV component, the corresponding values $-6.83$ and $11.0$ lie much farther into the tail.  The ratio $\Gamma_+/\Gamma_\parallel$ follows from these polarization-dependent projections and the root derivative in Eq.~(\ref{eq:resonant-damping}).

The radius follows from the transverse coupling of the two roots.  A root-centered Schur reduction gives, to leading order in $u_\perp$,
\begin{equation}
 (w_+-w_\parallel)^2+4u_\perp^2v_{\rm mix}^{2}=0,
 \qquad
 v_{\rm mix}^{2}=\frac{g_{+z}g_{z+}}{m_+'m_\parallel'},
\label{eq:discriminant}
\end{equation}
where primes denote frequency derivatives at the two axial roots and $g_{+z},g_{z+}$ are the transverse polarization-mixing matrix elements.  When their real frequencies have been tuned equal, $w_+-w_\parallel=-\ii\Delta\Gamma$, so Eq.~(\ref{eq:discriminant}) predicts
\begin{equation}
 \boxed{R_{\rm EP}^{\rm pred}=\frac{|\Delta\Gamma|}{2|v_{\rm mix}|}}
 \quad\text{to leading order.}
\label{eq:radius-law}
\end{equation}
The ring radius therefore records mode-selective microscopic damping.  At each parameter point, the axial roots fix $\Delta\Gamma$ and their root-normalized matrix elements fix $v_{\rm mix}$.  Both quantities are obtained before the off-axis double root is computed, and no ring-radius parameter is fitted.

For the baseline, $|v_{\rm mix}|=0.3012$ gives $R_{\rm EP}^{\rm pred}=2.4705\times10^{-3}$; the nonlinear double-root calculation gives $R_{\rm EP}^{\rm full}=2.4701\times10^{-3}$.  The parameter-free comparison covers 26 points obtained by varying temperature, $k_z$, $B_0$, density, and the temperature ratio.  Across these scans, $|\Delta\Gamma|$ changes by a factor of 310 and $R_{\rm EP}^{\rm full}$ by a factor of 343.  The ratio $R_{\rm EP}^{\rm full}/R_{\rm EP}^{\rm pred}$ remains within $0.12\%$ of unity [Fig.~\ref{fig:bulk}(b,c)].

Replacing $\Delta\Gamma$ by the mean damping $\bar\Gamma$ tests whether common attenuation can set the radius.  That proxy misses by $6$--$50\%$, as expected because common decay enters the two-mode identity component rather than its splitting.  The leading-order law applies to a small ring formed from two weakly damped, isolated roots.  Frequency-dependent self-energy, cone curvature, and complex mixing produce higher-order, parameter-dependent corrections [Fig.~\ref{fig:bulk}(d)].  Scan definitions and nonlinear-root checks are documented in the SM~\cite{SupplementalMaterial}.

The nonlinear conditions $\det\Dret=\partial_w\det\Dret=0$ locate the baseline ring at
\begin{equation}
\begin{aligned}
 w_{\rm EP}&=0.6178-8.1\times10^{-4}\ii,\\
 R_{\rm EP}^{\rm full}&=2.47\times10^{-3},\qquad X_{\rm EP}=0.222.
\end{aligned}
\label{eq:epnumbers}
\end{equation}
At the double root, $\Dret$ has a one-dimensional electric null space.  A linking loop exchanges the two nonlinear roots after one turn and restores them after two~\cite{SupplementalMaterial}.  This square-root monodromy and the geometric defectiveness identify an exceptional ring, rather than a broadened avoided crossing.  The same double-root structure persists throughout the five parameter scans.

\begin{figure*}[t]
 \includegraphics[width=\textwidth]{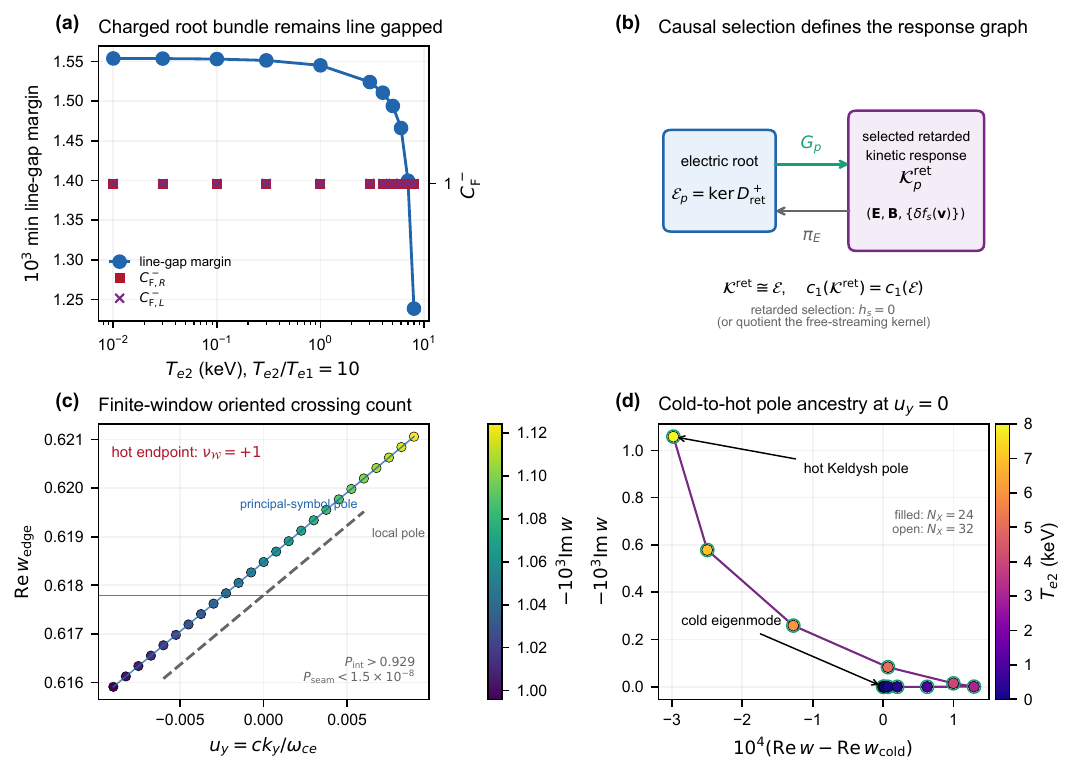}
 \caption{\label{fig:edge}The inherited charge and the localized Keldysh pole.  (a) Right and left electric root line bundles retain $C^-_{\rm F}=+1$ and a finite line gap during the cold--hot continuation.  (b) Causal reconstruction maps the electric root to the selected retarded kinetic response; $\delta f_s(\bm v)$ belongs to its infinite-dimensional fiber.  (c) Weyl quantization of the principal-symbol response gives a localized Keldysh pole with finite-window oriented crossing count $\nu_{\mathcal W}=+1$.  Color denotes damping.  (d) At $u_y=0$, temperature continuation connects the cold Hermitian interface mode to the hot simple pole.}
\end{figure*}

\emph{Causal selection defines the kinetic-response bundle.---}
Having established how kinetic resonances reshape the degeneracy, we next ask whether eliminating velocity space preserves its charge.  An electric root does not by itself specify a kinetic response because the Vlasov equation also admits homogeneous free-streaming solutions.  This obstruction is absent from finite-dimensional phenomenological models and differs from ordinary auxiliary-state descriptions of dispersive media~\cite{RamanFan2010,Silveirinha2015}.  Let $p=(u_x,u_y,X)$ lie on a closed enclosing surface $\Sigma$.  A selected simple root $w(p)$ defines the electric root line bundle
\begin{equation}
 \mathcal E_p=\ker\Dret[w(p),p]\subset\mathbb C^3,
 \qquad z(p)=\omega_{ce}w(p).
\end{equation}
The retarded prescription selects the response driven by a field switched on in the past.  It sets the homogeneous term $h_s$ to zero, or equivalently works after quotienting the free-streaming kernel.  The selected retarded kinetic-response bundle is then the graph shown in Fig.~\ref{fig:edge}(b),
\begin{equation}
 G_p\bm E=\left(
 \bm E,\frac{\bm k(p)\times\bm E}{z(p)},
 \{\mathcal R_s^+\mathcal S_s\bm E\}_s\right),
 \qquad
 \mathcal K_p^{\rm ret}=G_p\mathcal E_p,
\label{eq:graph-map}
\end{equation}
where $\mathcal R_s^+$ is the retarded kinetic inverse and $\mathcal S_s$ the field source.  Electric projection is the inverse of $G_p$, so
\begin{equation}
 \mathcal K^{\rm ret}\cong\mathcal E,
 \qquad c_1(\mathcal K^{\rm ret})=c_1(\mathcal E).
\label{eq:bundle-isomorphism}
\end{equation}
For any continuous family $G_p$, projection from its graph to the electric component is a standard complex line-bundle isomorphism.  The kinetic content lies in the retarded prescription, which selects a single continuous family $G_p$ from the Vlasov continuum.  The construction requires a simple line-gap-isolated root, $z(p)\neq0$, and a continuous retarded response on $\Sigma$.  The calculation verifies root isolation.  For the Maxwellian sources used here, the SM places $\delta f_s$ in the strong dual of an analytic-Schwartz velocity space and establishes continuity on bounded electric-field sets with a single Landau contour below all resonance poles.  The same contour fixes the retarded sheet globally.  Quotienting removes the homogeneous $\bm E=0$ sector by definition, so no Chern class is assigned to the entire conservative Maxwell--Vlasov generator.  Because $G_p$ need not be unitary, electric and reconstructed Berry curvatures may differ pointwise, although their first Chern classes agree.  The full proposition is stated in the SM~\cite{SupplementalMaterial}.

\emph{The exceptional ring inherits the Weyl charge.---}
On an ellipsoid surrounding the ring, a vertical line through $\operatorname{Re}w_{\rm EP}$ separates the two selected roots.  The line gap refers only to this two-root sector on $\Sigma$, not to every root of the nonlinear pencil.  For normalized right electric null vectors $|E_i^R\rangle\in\mathcal E_{p_i}$, a gauge-invariant triangular link construction~\cite{Fukui2005}, related to simplicial characteristic-class methods~\cite{Bohlsen2026}, gives
\begin{equation}
 C^-_{\rm F}=-\frac{1}{2\pi}\sum_{(ijk)}
 \arg\!\left[
 \langle E_i^R|E_j^R\rangle\langle E_j^R|E_k^R\rangle
 \langle E_k^R|E_i^R\rangle\right]=+1.
\label{eq:chern}
\end{equation}
The sign follows Faure's outward orientation in $(u_x,u_y,X)$, including the decreasing density profile.  The left electric root line bundle gives the same integer.  The right and left electric root bundles remain line gapped and retain $C^-_{\rm F}=+1$ at all 11 sampled temperatures [Fig.~\ref{fig:edge}(a)].  Equation~(\ref{eq:bundle-isomorphism}) assigns the same class to the selected retarded kinetic-response bundle.  Along the computed continuation, damping redistributes the cold point singularity into a ring without changing the enclosed charge.  Mesh, orientation, and continuation checks are given in the SM~\cite{SupplementalMaterial}.

\emph{The cold interface mode continues as a localized pole.---}
At finite temperature the interface excitation is no longer a real-frequency eigenmode, so it must be followed as a pole of the retarded boundary response.  Weyl quantization~\cite{Zworski2012} of the nonlinear principal symbol gives
\begin{equation}
\begin{aligned}
 \mathsf T(w)&=\Op\Dret(w),\\[-2pt]
 \mathsf T_{ij}(w)
 &=\frac1{N_X}\sum_m e^{\ii q_m(X_i-X_j)}\\[-2pt]
 &\quad\times\Dret\!\left(w,\epsilon q_m,u_y,
 \frac{X_i+X_j}{2}\right),\\
 \epsilon&=\frac{c}{\omega_{ce}L}.
\end{aligned}
\label{eq:weylquant}
\end{equation}
A local expansion near the WER predicts a Gaussian pole with $w_{\rm loc}=0.6178-1.04\times10^{-3}\ii$, rms width $0.133L$, and slope $\partial w/\partial u_y\simeq0.286-0.007\ii$.  Its localization samples momenta beyond the linear neighborhood of the ring.  Equation~(\ref{eq:weylquant}) therefore retains the frequency and $u_x$ dependence of the retarded symbol, with explicit operator ordering for subprincipal tests; the local solution and discretization are derived in the SM~\cite{SupplementalMaterial}.

A response pole occurs when this operator has a zero eigenvalue with nonzero Keldysh denominator.  At $u_y=0$, the calculation converges to
\begin{equation}
 w_{\rm edge}=0.6185-1.06\times10^{-3}\ii,
 \qquad \sigma_X=0.135L.
\label{eq:edgepole}
\end{equation}
The nonzero denominator $\kappa=\langle L|\partial_w(\Op\Dret)|R\rangle$ makes this a rank-one simple Keldysh pole.  Its electric field is concentrated at the physical interface.  A 12-node temperature continuation at $u_y=0$ connects the cold Hermitian mode at $w=0.61878$ to this pole without a state switch [Fig.~\ref{fig:edge}(d)].  The continuation identifies the hot pole as the descendant of the cold interface mode.  Fourier, box-length, ordering, ion, and profile-width checks are reported in the SM~\cite{SupplementalMaterial}.

At the hot endpoint, a 25-point momentum scan over $\mathcal W=[-0.009,0.009]$ gives one upward crossing of $\operatorname{Re}w=\operatorname{Re}w_{\rm EP}$ and no downward crossing.  We define the finite-window oriented crossing count as
\begin{equation}
 \nu_{\mathcal W}=N_\uparrow-N_\downarrow=+1=C^-_{\rm F}.
\label{eq:endpoint-match}
\end{equation}
The crossing occurs at $u_y=-2.42\times10^{-3}$, with slope $\partial w/\partial u_y=0.2857-0.00713\ii$ [Fig.~\ref{fig:edge}(c)].  Thus the real frequency crosses the reference line while the damping changes weakly.  This comparison concerns the line-gap Chern charge, not exceptional-ring vorticity or point-gap winding.

A sampled contour search finds no competing crossing within the stated frequency ellipse and momentum window.  Together with the temperature continuation, the hot scan provides finite-window evidence consistent with extending the cold phase-space index relation to the selected nonlinear retarded response.  It does not establish a global kinetic index theorem.

Relativistic and finite-orbit calculations shift the ring and pole quantitatively while preserving the computed charge, localization, and crossing direction~\cite{Juttner1911,YoonDavidson1990,SupplementalMaterial}.  The relativistic point is not included in the nonrelativistic radius-law collapse, and these tests do not define a Chern class for the nontranslation-invariant relativistic operator.

For $B_0=1\,{\rm T}$, the orbit-resolved pole corresponds to about $17\,{\rm GHz}$, a $2\,{\rm ns}$ amplitude-decay time, a $1.4\,{\rm cm}$ localization length, and a $0.2\,{\rm m}$ damping length.  This mechanism differs from the continuum damping studied by Rajawat \emph{et al.}, where a cold surface-plasma wave converts into upper-hybrid modes across a smooth boundary~\cite{Rajawat2025}.  Here the imaginary self-energy is already present in the locally homogeneous Vlasov response and creates the WER before the interface problem is solved.

Mode-selective collisionless resonances set the exceptional geometry: differential damping opens the ring, while the projected self-energy sets the pole lifetime.  Under the stated spectral assumptions, retarded graph reconstruction identifies a causally selected kinetic response that carries the electric-root Chern class.  At the hot endpoint, its localized pole has a finite-window oriented crossing count consistent with the inherited charge.  A general index theorem for nonlinear retarded pencils still requires control of the full parameter homotopy and additional continuum or purely kinetic branches.  Within these limits, the calculation identifies conditions under which integrating out a conservative continuum preserves the topology of the causally selected response.

\begin{acknowledgments}
The authors acknowledge Xeonova Ltd. for providing computational resources and technical support. The authors also thank their colleagues for helpful discussions.
\end{acknowledgments}

\paragraph*{Data and code availability.}
The source code, machine-readable root and continuation data, figure scripts, and exact dependency lock supporting this work are archived in Zenodo~\cite{RaoZenodo2026}.  Source code is released under the BSD 3-Clause License and research data under CC BY 4.0.  The SM lists the reproducible commands and the scope of each numerical certificate.

\bibliography{references}

\begin{thebibliography}{44}%
\makeatletter
\providecommand \@ifxundefined [1]{%
 \@ifx{#1\undefined}
}%
\providecommand \@ifnum [1]{%
 \ifnum #1\expandafter \@firstoftwo
 \else \expandafter \@secondoftwo
 \fi
}%
\providecommand \@ifx [1]{%
 \ifx #1\expandafter \@firstoftwo
 \else \expandafter \@secondoftwo
 \fi
}%
\providecommand \natexlab [1]{#1}%
\providecommand \enquote  [1]{``#1''}%
\providecommand \bibnamefont  [1]{#1}%
\providecommand \bibfnamefont [1]{#1}%
\providecommand \citenamefont [1]{#1}%
\providecommand \href@noop [0]{\@secondoftwo}%
\providecommand \href [0]{\begingroup \@sanitize@url \@href}%
\providecommand \@href[1]{\@@startlink{#1}\@@href}%
\providecommand \@@href[1]{\endgroup#1\@@endlink}%
\providecommand \@sanitize@url [0]{\catcode `\\12\catcode `\$12\catcode
  `\&12\catcode `\#12\catcode `\^12\catcode `\_12\catcode `\%12\relax}%
\providecommand \@@startlink[1]{}%
\providecommand \@@endlink[0]{}%
\providecommand \url  [0]{\begingroup\@sanitize@url \@url }%
\providecommand \@url [1]{\endgroup\@href {#1}{\urlprefix }}%
\providecommand \urlprefix  [0]{URL }%
\providecommand \Eprint [0]{\href }%
\providecommand \doibase [0]{https://doi.org/}%
\providecommand \selectlanguage [0]{\@gobble}%
\providecommand \bibinfo  [0]{\@secondoftwo}%
\providecommand \bibfield  [0]{\@secondoftwo}%
\providecommand \translation [1]{[#1]}%
\providecommand \BibitemOpen [0]{}%
\providecommand \bibitemStop [0]{}%
\providecommand \bibitemNoStop [0]{.\EOS\space}%
\providecommand \EOS [0]{\spacefactor3000\relax}%
\providecommand \BibitemShut  [1]{\csname bibitem#1\endcsname}%
\let\auto@bib@innerbib\@empty
\bibitem [{\citenamefont {Raman}\ and\ \citenamefont
  {Fan}(2010)}]{RamanFan2010}%
  \BibitemOpen
  \bibfield  {author} {\bibinfo {author} {\bibfnamefont {A.}~\bibnamefont
  {Raman}}\ and\ \bibinfo {author} {\bibfnamefont {S.}~\bibnamefont {Fan}},\
  }\bibfield  {title} {\bibinfo {title} {Photonic band structure of dispersive
  metamaterials formulated as a hermitian eigenvalue problem},\ }\href
  {https://doi.org/10.1103/PhysRevLett.104.087401} {\bibfield  {journal}
  {\bibinfo  {journal} {Phys. Rev. Lett.}\ }\textbf {\bibinfo {volume} {104}},\
  \bibinfo {pages} {087401} (\bibinfo {year} {2010})}\BibitemShut {NoStop}%
\bibitem [{\citenamefont {Silveirinha}(2015)}]{Silveirinha2015}%
  \BibitemOpen
  \bibfield  {author} {\bibinfo {author} {\bibfnamefont {M.~G.}\ \bibnamefont
  {Silveirinha}},\ }\bibfield  {title} {\bibinfo {title} {Chern invariants for
  continuous media},\ }\href {https://doi.org/10.1103/PhysRevB.92.125153}
  {\bibfield  {journal} {\bibinfo  {journal} {Phys. Rev. B}\ }\textbf {\bibinfo
  {volume} {92}},\ \bibinfo {pages} {125153} (\bibinfo {year}
  {2015})}\BibitemShut {NoStop}%
\bibitem [{\citenamefont {Zhen}\ \emph {et~al.}(2015)\citenamefont {Zhen},
  \citenamefont {Hsu}, \citenamefont {Igarashi}, \citenamefont {Lu},
  \citenamefont {Kaminer}, \citenamefont {Pick}, \citenamefont {Chua},
  \citenamefont {Joannopoulos},\ and\ \citenamefont
  {Solja\v{c}i\'c}}]{Zhen2015}%
  \BibitemOpen
  \bibfield  {author} {\bibinfo {author} {\bibfnamefont {B.}~\bibnamefont
  {Zhen}}, \bibinfo {author} {\bibfnamefont {C.~W.}\ \bibnamefont {Hsu}},
  \bibinfo {author} {\bibfnamefont {Y.}~\bibnamefont {Igarashi}}, \bibinfo
  {author} {\bibfnamefont {L.}~\bibnamefont {Lu}}, \bibinfo {author}
  {\bibfnamefont {I.}~\bibnamefont {Kaminer}}, \bibinfo {author} {\bibfnamefont
  {A.}~\bibnamefont {Pick}}, \bibinfo {author} {\bibfnamefont {S.-L.}\
  \bibnamefont {Chua}}, \bibinfo {author} {\bibfnamefont {J.~D.}\ \bibnamefont
  {Joannopoulos}},\ and\ \bibinfo {author} {\bibfnamefont {M.}~\bibnamefont
  {Solja\v{c}i\'c}},\ }\bibfield  {title} {\bibinfo {title} {Spawning rings of
  exceptional points out of {Dirac} cones},\ }\href
  {https://doi.org/10.1038/nature14889} {\bibfield  {journal} {\bibinfo
  {journal} {Nature}\ }\textbf {\bibinfo {volume} {525}},\ \bibinfo {pages}
  {354} (\bibinfo {year} {2015})}\BibitemShut {NoStop}%
\bibitem [{\citenamefont {Xu}\ \emph {et~al.}(2017)\citenamefont {Xu},
  \citenamefont {Wang},\ and\ \citenamefont {Duan}}]{Xu2017}%
  \BibitemOpen
  \bibfield  {author} {\bibinfo {author} {\bibfnamefont {Y.}~\bibnamefont
  {Xu}}, \bibinfo {author} {\bibfnamefont {S.-T.}\ \bibnamefont {Wang}},\ and\
  \bibinfo {author} {\bibfnamefont {L.-M.}\ \bibnamefont {Duan}},\ }\bibfield
  {title} {\bibinfo {title} {Weyl exceptional rings in a three-dimensional
  dissipative cold atomic gas},\ }\href
  {https://doi.org/10.1103/PhysRevLett.118.045701} {\bibfield  {journal}
  {\bibinfo  {journal} {Phys. Rev. Lett.}\ }\textbf {\bibinfo {volume} {118}},\
  \bibinfo {pages} {045701} (\bibinfo {year} {2017})}\BibitemShut {NoStop}%
\bibitem [{\citenamefont {Cerjan}\ \emph {et~al.}(2019)\citenamefont {Cerjan},
  \citenamefont {Huang}, \citenamefont {Wang}, \citenamefont {Chen},
  \citenamefont {Chong},\ and\ \citenamefont {Rechtsman}}]{Cerjan2019}%
  \BibitemOpen
  \bibfield  {author} {\bibinfo {author} {\bibfnamefont {A.}~\bibnamefont
  {Cerjan}}, \bibinfo {author} {\bibfnamefont {S.}~\bibnamefont {Huang}},
  \bibinfo {author} {\bibfnamefont {M.}~\bibnamefont {Wang}}, \bibinfo {author}
  {\bibfnamefont {K.~P.}\ \bibnamefont {Chen}}, \bibinfo {author}
  {\bibfnamefont {Y.}~\bibnamefont {Chong}},\ and\ \bibinfo {author}
  {\bibfnamefont {M.~C.}\ \bibnamefont {Rechtsman}},\ }\bibfield  {title}
  {\bibinfo {title} {Experimental realization of a weyl exceptional ring},\
  }\href {https://doi.org/10.1038/s41566-019-0453-z} {\bibfield  {journal}
  {\bibinfo  {journal} {Nat. Photonics}\ }\textbf {\bibinfo {volume} {13}},\
  \bibinfo {pages} {623} (\bibinfo {year} {2019})}\BibitemShut {NoStop}%
\bibitem [{\citenamefont {Mc~Guinness}\ and\ \citenamefont
  {Eastham}(2020)}]{McGuinness2020}%
  \BibitemOpen
  \bibfield  {author} {\bibinfo {author} {\bibfnamefont {R.~L.}\ \bibnamefont
  {Mc~Guinness}}\ and\ \bibinfo {author} {\bibfnamefont {P.~R.}\ \bibnamefont
  {Eastham}},\ }\bibfield  {title} {\bibinfo {title} {Weyl points and
  exceptional rings with polaritons in bulk semiconductors},\ }\href
  {https://doi.org/10.1103/PhysRevResearch.2.043268} {\bibfield  {journal}
  {\bibinfo  {journal} {Phys. Rev. Research}\ }\textbf {\bibinfo {volume}
  {2}},\ \bibinfo {pages} {043268} (\bibinfo {year} {2020})}\BibitemShut
  {NoStop}%
\bibitem [{\citenamefont {Bergholtz}\ \emph {et~al.}(2021)\citenamefont
  {Bergholtz}, \citenamefont {Budich},\ and\ \citenamefont
  {Kunst}}]{Bergholtz2021}%
  \BibitemOpen
  \bibfield  {author} {\bibinfo {author} {\bibfnamefont {E.~J.}\ \bibnamefont
  {Bergholtz}}, \bibinfo {author} {\bibfnamefont {J.~C.}\ \bibnamefont
  {Budich}},\ and\ \bibinfo {author} {\bibfnamefont {F.~K.}\ \bibnamefont
  {Kunst}},\ }\bibfield  {title} {\bibinfo {title} {Exceptional topology of
  non-hermitian systems},\ }\href
  {https://doi.org/10.1103/RevModPhys.93.015005} {\bibfield  {journal}
  {\bibinfo  {journal} {Rev. Mod. Phys.}\ }\textbf {\bibinfo {volume} {93}},\
  \bibinfo {pages} {015005} (\bibinfo {year} {2021})}\BibitemShut {NoStop}%
\bibitem [{\citenamefont {Kawabata}\ \emph {et~al.}(2019)\citenamefont
  {Kawabata}, \citenamefont {Shiozaki}, \citenamefont {Ueda},\ and\
  \citenamefont {Sato}}]{Kawabata2019}%
  \BibitemOpen
  \bibfield  {author} {\bibinfo {author} {\bibfnamefont {K.}~\bibnamefont
  {Kawabata}}, \bibinfo {author} {\bibfnamefont {K.}~\bibnamefont {Shiozaki}},
  \bibinfo {author} {\bibfnamefont {M.}~\bibnamefont {Ueda}},\ and\ \bibinfo
  {author} {\bibfnamefont {M.}~\bibnamefont {Sato}},\ }\bibfield  {title}
  {\bibinfo {title} {Symmetry and topology in non-hermitian physics},\ }\href
  {https://doi.org/10.1103/PhysRevX.9.041015} {\bibfield  {journal} {\bibinfo
  {journal} {Phys. Rev. X}\ }\textbf {\bibinfo {volume} {9}},\ \bibinfo {pages}
  {041015} (\bibinfo {year} {2019})}\BibitemShut {NoStop}%
\bibitem [{\citenamefont {Kotz}\ and\ \citenamefont
  {Timm}(2023)}]{KotzTimm2023}%
  \BibitemOpen
  \bibfield  {author} {\bibinfo {author} {\bibfnamefont {M.}~\bibnamefont
  {Kotz}}\ and\ \bibinfo {author} {\bibfnamefont {C.}~\bibnamefont {Timm}},\
  }\bibfield  {title} {\bibinfo {title} {Topological classification of
  non-hermitian hamiltonians with frequency dependence},\ }\href
  {https://doi.org/10.1103/PhysRevResearch.5.033043} {\bibfield  {journal}
  {\bibinfo  {journal} {Phys. Rev. Research}\ }\textbf {\bibinfo {volume}
  {5}},\ \bibinfo {pages} {033043} (\bibinfo {year} {2023})}\BibitemShut
  {NoStop}%
\bibitem [{\citenamefont {Isobe}\ \emph {et~al.}(2024)\citenamefont {Isobe},
  \citenamefont {Yoshida},\ and\ \citenamefont {Hatsugai}}]{Isobe2024}%
  \BibitemOpen
  \bibfield  {author} {\bibinfo {author} {\bibfnamefont {T.}~\bibnamefont
  {Isobe}}, \bibinfo {author} {\bibfnamefont {T.}~\bibnamefont {Yoshida}},\
  and\ \bibinfo {author} {\bibfnamefont {Y.}~\bibnamefont {Hatsugai}},\
  }\bibfield  {title} {\bibinfo {title} {Bulk-edge correspondence for nonlinear
  eigenvalue problems},\ }\href
  {https://doi.org/10.1103/PhysRevLett.132.126601} {\bibfield  {journal}
  {\bibinfo  {journal} {Phys. Rev. Lett.}\ }\textbf {\bibinfo {volume} {132}},\
  \bibinfo {pages} {126601} (\bibinfo {year} {2024})}\BibitemShut {NoStop}%
\bibitem [{\citenamefont {Yoshida}\ \emph {et~al.}(2025)\citenamefont
  {Yoshida}, \citenamefont {Isobe},\ and\ \citenamefont
  {Hatsugai}}]{Yoshida2025}%
  \BibitemOpen
  \bibfield  {author} {\bibinfo {author} {\bibfnamefont {T.}~\bibnamefont
  {Yoshida}}, \bibinfo {author} {\bibfnamefont {T.}~\bibnamefont {Isobe}},\
  and\ \bibinfo {author} {\bibfnamefont {Y.}~\bibnamefont {Hatsugai}},\
  }\bibfield  {title} {\bibinfo {title} {Exceptional points and non-hermitian
  skin effects under nonlinearity of eigenvalues},\ }\href
  {https://doi.org/10.1103/PhysRevB.111.064310} {\bibfield  {journal} {\bibinfo
   {journal} {Phys. Rev. B}\ }\textbf {\bibinfo {volume} {111}},\ \bibinfo
  {pages} {064310} (\bibinfo {year} {2025})}\BibitemShut {NoStop}%
\bibitem [{\citenamefont {Delplace}\ \emph {et~al.}(2017)\citenamefont
  {Delplace}, \citenamefont {Marston},\ and\ \citenamefont
  {Venaille}}]{Delplace2017}%
  \BibitemOpen
  \bibfield  {author} {\bibinfo {author} {\bibfnamefont {P.}~\bibnamefont
  {Delplace}}, \bibinfo {author} {\bibfnamefont {J.~B.}\ \bibnamefont
  {Marston}},\ and\ \bibinfo {author} {\bibfnamefont {A.}~\bibnamefont
  {Venaille}},\ }\bibfield  {title} {\bibinfo {title} {Topological origin of
  equatorial waves},\ }\href {https://doi.org/10.1126/science.aan8819}
  {\bibfield  {journal} {\bibinfo  {journal} {Science}\ }\textbf {\bibinfo
  {volume} {358}},\ \bibinfo {pages} {1075} (\bibinfo {year}
  {2017})}\BibitemShut {NoStop}%
\bibitem [{\citenamefont {Faure}(2023)}]{Faure2023}%
  \BibitemOpen
  \bibfield  {author} {\bibinfo {author} {\bibfnamefont {F.}~\bibnamefont
  {Faure}},\ }\bibfield  {title} {\bibinfo {title} {Manifestation of the
  topological index formula in quantum waves and geophysical waves},\ }\href
  {https://doi.org/10.5802/ahl.169} {\bibfield  {journal} {\bibinfo  {journal}
  {Ann. Henri Lebesgue}\ }\textbf {\bibinfo {volume} {6}},\ \bibinfo {pages}
  {449} (\bibinfo {year} {2023})}\BibitemShut {NoStop}%
\bibitem [{\citenamefont {Qin}\ and\ \citenamefont {Fu}(2023)}]{QinFu2023}%
  \BibitemOpen
  \bibfield  {author} {\bibinfo {author} {\bibfnamefont {H.}~\bibnamefont
  {Qin}}\ and\ \bibinfo {author} {\bibfnamefont {Y.}~\bibnamefont {Fu}},\
  }\bibfield  {title} {\bibinfo {title} {Topological langmuir-cyclotron wave},\
  }\href {https://doi.org/10.1126/sciadv.add8041} {\bibfield  {journal}
  {\bibinfo  {journal} {Sci. Adv.}\ }\textbf {\bibinfo {volume} {9}},\ \bibinfo
  {pages} {eadd8041} (\bibinfo {year} {2023})}\BibitemShut {NoStop}%
\bibitem [{\citenamefont {Jezequel}\ and\ \citenamefont
  {Delplace}(2023)}]{Jezequel2023}%
  \BibitemOpen
  \bibfield  {author} {\bibinfo {author} {\bibfnamefont {L.}~\bibnamefont
  {Jezequel}}\ and\ \bibinfo {author} {\bibfnamefont {P.}~\bibnamefont
  {Delplace}},\ }\bibfield  {title} {\bibinfo {title} {Non-hermitian spectral
  flows and berry--chern monopoles},\ }\href
  {https://doi.org/10.1103/PhysRevLett.130.066601} {\bibfield  {journal}
  {\bibinfo  {journal} {Phys. Rev. Lett.}\ }\textbf {\bibinfo {volume} {130}},\
  \bibinfo {pages} {066601} (\bibinfo {year} {2023})}\BibitemShut {NoStop}%
\bibitem [{\citenamefont {Morrison}(1980)}]{Morrison1980}%
  \BibitemOpen
  \bibfield  {author} {\bibinfo {author} {\bibfnamefont {P.~J.}\ \bibnamefont
  {Morrison}},\ }\bibfield  {title} {\bibinfo {title} {The {Maxwell--Vlasov}
  equations as a continuous {Hamiltonian} system},\ }\href
  {https://doi.org/10.1016/0375-9601(80)90776-8} {\bibfield  {journal}
  {\bibinfo  {journal} {Phys. Lett. A}\ }\textbf {\bibinfo {volume} {80}},\
  \bibinfo {pages} {383} (\bibinfo {year} {1980})}\BibitemShut {NoStop}%
\bibitem [{\citenamefont {van Kampen}(1955)}]{VanKampen1955}%
  \BibitemOpen
  \bibfield  {author} {\bibinfo {author} {\bibfnamefont {N.~G.}\ \bibnamefont
  {van Kampen}},\ }\bibfield  {title} {\bibinfo {title} {On the theory of
  stationary waves in plasmas},\ }\href
  {https://doi.org/10.1016/S0031-8914(55)93068-8} {\bibfield  {journal}
  {\bibinfo  {journal} {Physica}\ }\textbf {\bibinfo {volume} {21}},\ \bibinfo
  {pages} {949} (\bibinfo {year} {1955})}\BibitemShut {NoStop}%
\bibitem [{\citenamefont {Case}(1959)}]{Case1959}%
  \BibitemOpen
  \bibfield  {author} {\bibinfo {author} {\bibfnamefont {K.~M.}\ \bibnamefont
  {Case}},\ }\bibfield  {title} {\bibinfo {title} {Plasma oscillations},\
  }\href {https://doi.org/10.1016/0003-4916(59)90029-6} {\bibfield  {journal}
  {\bibinfo  {journal} {Ann. Phys.}\ }\textbf {\bibinfo {volume} {7}},\
  \bibinfo {pages} {349} (\bibinfo {year} {1959})}\BibitemShut {NoStop}%
\bibitem [{\citenamefont {Ramos}(2019)}]{Ramos2019}%
  \BibitemOpen
  \bibfield  {author} {\bibinfo {author} {\bibfnamefont {J.~J.}\ \bibnamefont
  {Ramos}},\ }\bibfield  {title} {\bibinfo {title} {Normal-mode-based theory of
  collisionless plasma waves},\ }\href
  {https://doi.org/10.1017/S0022377819000400} {\bibfield  {journal} {\bibinfo
  {journal} {J. Plasma Phys.}\ }\textbf {\bibinfo {volume} {85}},\ \bibinfo
  {pages} {905850401} (\bibinfo {year} {2019})}\BibitemShut {NoStop}%
\bibitem [{\citenamefont {Huang}\ \emph {et~al.}(2026)\citenamefont {Huang},
  \citenamefont {Hu},\ and\ \citenamefont {Yang}}]{Huang2026}%
  \BibitemOpen
  \bibfield  {author} {\bibinfo {author} {\bibfnamefont {J.}~\bibnamefont
  {Huang}}, \bibinfo {author} {\bibfnamefont {J.}~\bibnamefont {Hu}},\ and\
  \bibinfo {author} {\bibfnamefont {Z.}~\bibnamefont {Yang}},\ }\bibfield
  {title} {\bibinfo {title} {Complex frequency detection in a subsystem},\
  }\href {https://doi.org/10.1038/s42005-026-02524-8} {\bibfield  {journal}
  {\bibinfo  {journal} {Commun. Phys.}\ }\textbf {\bibinfo {volume} {9}},\
  \bibinfo {pages} {84} (\bibinfo {year} {2026})}\BibitemShut {NoStop}%
\bibitem [{\citenamefont {Kammerer}\ \emph {et~al.}(2008)\citenamefont
  {Kammerer}, \citenamefont {Merz},\ and\ \citenamefont
  {Jenko}}]{Kammerer2008}%
  \BibitemOpen
  \bibfield  {author} {\bibinfo {author} {\bibfnamefont {M.}~\bibnamefont
  {Kammerer}}, \bibinfo {author} {\bibfnamefont {F.}~\bibnamefont {Merz}},\
  and\ \bibinfo {author} {\bibfnamefont {F.}~\bibnamefont {Jenko}},\ }\bibfield
   {title} {\bibinfo {title} {Exceptional points in linear gyrokinetics},\
  }\href {https://doi.org/10.1063/1.2909618} {\bibfield  {journal} {\bibinfo
  {journal} {Phys. Plasmas}\ }\textbf {\bibinfo {volume} {15}},\ \bibinfo
  {pages} {052102} (\bibinfo {year} {2008})}\BibitemShut {NoStop}%
\bibitem [{\citenamefont {Gao}\ \emph {et~al.}(2016)\citenamefont {Gao},
  \citenamefont {Yang}, \citenamefont {Lawrence}, \citenamefont {Fang},
  \citenamefont {B\'eri},\ and\ \citenamefont {Zhang}}]{Gao2016}%
  \BibitemOpen
  \bibfield  {author} {\bibinfo {author} {\bibfnamefont {W.}~\bibnamefont
  {Gao}}, \bibinfo {author} {\bibfnamefont {B.}~\bibnamefont {Yang}}, \bibinfo
  {author} {\bibfnamefont {M.}~\bibnamefont {Lawrence}}, \bibinfo {author}
  {\bibfnamefont {F.}~\bibnamefont {Fang}}, \bibinfo {author} {\bibfnamefont
  {B.}~\bibnamefont {B\'eri}},\ and\ \bibinfo {author} {\bibfnamefont
  {S.}~\bibnamefont {Zhang}},\ }\bibfield  {title} {\bibinfo {title} {Photonic
  weyl degeneracies in magnetized plasma},\ }\href
  {https://doi.org/10.1038/ncomms12435} {\bibfield  {journal} {\bibinfo
  {journal} {Nat. Commun.}\ }\textbf {\bibinfo {volume} {7}},\ \bibinfo {pages}
  {12435} (\bibinfo {year} {2016})}\BibitemShut {NoStop}%
\bibitem [{\citenamefont {Fu}\ and\ \citenamefont {Qin}(2021)}]{FuQin2021}%
  \BibitemOpen
  \bibfield  {author} {\bibinfo {author} {\bibfnamefont {Y.}~\bibnamefont
  {Fu}}\ and\ \bibinfo {author} {\bibfnamefont {H.}~\bibnamefont {Qin}},\
  }\bibfield  {title} {\bibinfo {title} {Topological phases and bulk-edge
  correspondence of magnetized cold plasmas},\ }\href
  {https://doi.org/10.1038/s41467-021-24189-3} {\bibfield  {journal} {\bibinfo
  {journal} {Nat. Commun.}\ }\textbf {\bibinfo {volume} {12}},\ \bibinfo
  {pages} {3924} (\bibinfo {year} {2021})}\BibitemShut {NoStop}%
\bibitem [{\citenamefont {Fu}\ and\ \citenamefont {Qin}(2022)}]{FuQin2022}%
  \BibitemOpen
  \bibfield  {author} {\bibinfo {author} {\bibfnamefont {Y.}~\bibnamefont
  {Fu}}\ and\ \bibinfo {author} {\bibfnamefont {H.}~\bibnamefont {Qin}},\
  }\bibfield  {title} {\bibinfo {title} {The dispersion and propagation of
  topological {Langmuir-cyclotron} waves in cold magnetized plasmas},\ }\href
  {https://doi.org/10.1017/S0022377822000629} {\bibfield  {journal} {\bibinfo
  {journal} {J. Plasma Phys.}\ }\textbf {\bibinfo {volume} {88}},\ \bibinfo
  {pages} {835880401} (\bibinfo {year} {2022})}\BibitemShut {NoStop}%
\bibitem [{\citenamefont {Fonseca}\ \emph {et~al.}(2024)\citenamefont
  {Fonseca}, \citenamefont {Prud\^encio}, \citenamefont {Silveirinha},\ and\
  \citenamefont {Huidobro}}]{Fonseca2024}%
  \BibitemOpen
  \bibfield  {author} {\bibinfo {author} {\bibfnamefont {G.~R.}\ \bibnamefont
  {Fonseca}}, \bibinfo {author} {\bibfnamefont {F.~R.}\ \bibnamefont
  {Prud\^encio}}, \bibinfo {author} {\bibfnamefont {M.~G.}\ \bibnamefont
  {Silveirinha}},\ and\ \bibinfo {author} {\bibfnamefont {P.~A.}\ \bibnamefont
  {Huidobro}},\ }\bibfield  {title} {\bibinfo {title} {First-principles study
  of topological invariants of {Weyl} points in continuous media},\ }\href
  {https://doi.org/10.1103/PhysRevResearch.6.013017} {\bibfield  {journal}
  {\bibinfo  {journal} {Phys. Rev. Research}\ }\textbf {\bibinfo {volume}
  {6}},\ \bibinfo {pages} {013017} (\bibinfo {year} {2024})}\BibitemShut
  {NoStop}%
\bibitem [{\citenamefont {Rao}\ \emph {et~al.}(2025)\citenamefont {Rao},
  \citenamefont {Yolbarsop}, \citenamefont {Li},\ and\ \citenamefont
  {Liu}}]{Rao2025}%
  \BibitemOpen
  \bibfield  {author} {\bibinfo {author} {\bibfnamefont {X.}~\bibnamefont
  {Rao}}, \bibinfo {author} {\bibfnamefont {A.}~\bibnamefont {Yolbarsop}},
  \bibinfo {author} {\bibfnamefont {H.}~\bibnamefont {Li}},\ and\ \bibinfo
  {author} {\bibfnamefont {W.}~\bibnamefont {Liu}},\ }\bibfield  {title}
  {\bibinfo {title} {Symmetry-driven bulk-edge correspondence in electron
  magnetofluids at finite temperature},\ }\href
  {https://doi.org/10.1103/lvkj-3gdl} {\bibfield  {journal} {\bibinfo
  {journal} {Phys. Rev. Research}\ }\textbf {\bibinfo {volume} {7}},\ \bibinfo
  {pages} {023315} (\bibinfo {year} {2025})}\BibitemShut {NoStop}%
\bibitem [{\citenamefont {Billings}\ \emph {et~al.}(2026)\citenamefont
  {Billings}, \citenamefont {Qin}, \citenamefont {Ren},\ and\ \citenamefont
  {Marston}}]{Billings2026}%
  \BibitemOpen
  \bibfield  {author} {\bibinfo {author} {\bibfnamefont {V.}~\bibnamefont
  {Billings}}, \bibinfo {author} {\bibfnamefont {H.}~\bibnamefont {Qin}},
  \bibinfo {author} {\bibfnamefont {C.}~\bibnamefont {Ren}},\ and\ \bibinfo
  {author} {\bibfnamefont {J.~B.}\ \bibnamefont {Marston}},\ }\href
  {https://doi.org/10.48550/arXiv.2605.07845} {\bibinfo {title} {Warm
  topological langmuir cyclotron wave}} (\bibinfo {year} {2026}),\ \bibinfo
  {note} {preprint},\ \Eprint {https://arxiv.org/abs/2605.07845}
  {arXiv:2605.07845 [physics.plasm-ph]} \BibitemShut {NoStop}%
\bibitem [{\citenamefont {Fu}\ and\ \citenamefont {Qin}(2024)}]{FuQin2024}%
  \BibitemOpen
  \bibfield  {author} {\bibinfo {author} {\bibfnamefont {Y.}~\bibnamefont
  {Fu}}\ and\ \bibinfo {author} {\bibfnamefont {H.}~\bibnamefont {Qin}},\
  }\bibfield  {title} {\bibinfo {title} {Topological modes and spectral flows
  in inhomogeneous {$\mathcal{PT}$}-symmetric continuous media},\ }\href
  {https://doi.org/10.1103/PhysRevResearch.6.023273} {\bibfield  {journal}
  {\bibinfo  {journal} {Phys. Rev. Research}\ }\textbf {\bibinfo {volume}
  {6}},\ \bibinfo {pages} {023273} (\bibinfo {year} {2024})}\BibitemShut
  {NoStop}%
\bibitem [{\citenamefont {Rao}\ \emph {et~al.}(2026{\natexlab{a}})\citenamefont
  {Rao}, \citenamefont {Yolbarsop}, \citenamefont {Li},\ and\ \citenamefont
  {Liu}}]{Rao2026}%
  \BibitemOpen
  \bibfield  {author} {\bibinfo {author} {\bibfnamefont {X.}~\bibnamefont
  {Rao}}, \bibinfo {author} {\bibfnamefont {A.}~\bibnamefont {Yolbarsop}},
  \bibinfo {author} {\bibfnamefont {H.}~\bibnamefont {Li}},\ and\ \bibinfo
  {author} {\bibfnamefont {W.}~\bibnamefont {Liu}},\ }\bibfield  {title}
  {\bibinfo {title} {Phase-space topology and spectral flow in screened
  magnetized plasmas},\ }\href {https://doi.org/10.1103/kfhx-hp37} {\bibfield
  {journal} {\bibinfo  {journal} {Phys. Rev. Research}\ }\textbf {\bibinfo
  {volume} {8}},\ \bibinfo {pages} {023334} (\bibinfo {year}
  {2026}{\natexlab{a}})}\BibitemShut {NoStop}%
\bibitem [{\citenamefont {Wang}\ \emph {et~al.}(2020)\citenamefont {Wang},
  \citenamefont {Gao}, \citenamefont {Cao}, \citenamefont {Xiang},\ and\
  \citenamefont {Zhang}}]{Wang2020}%
  \BibitemOpen
  \bibfield  {author} {\bibinfo {author} {\bibfnamefont {W.}~\bibnamefont
  {Wang}}, \bibinfo {author} {\bibfnamefont {W.}~\bibnamefont {Gao}}, \bibinfo
  {author} {\bibfnamefont {L.}~\bibnamefont {Cao}}, \bibinfo {author}
  {\bibfnamefont {Y.}~\bibnamefont {Xiang}},\ and\ \bibinfo {author}
  {\bibfnamefont {S.}~\bibnamefont {Zhang}},\ }\bibfield  {title} {\bibinfo
  {title} {Photonic topological fermi nodal disk in non-hermitian magnetic
  plasma},\ }\href {https://doi.org/10.1038/s41377-020-0274-3} {\bibfield
  {journal} {\bibinfo  {journal} {Light Sci. Appl.}\ }\textbf {\bibinfo
  {volume} {9}},\ \bibinfo {pages} {40} (\bibinfo {year} {2020})}\BibitemShut
  {NoStop}%
\bibitem [{\citenamefont {Shastri}\ and\ \citenamefont
  {Monticone}(2020)}]{Shastri2020}%
  \BibitemOpen
  \bibfield  {author} {\bibinfo {author} {\bibfnamefont {K.}~\bibnamefont
  {Shastri}}\ and\ \bibinfo {author} {\bibfnamefont {F.}~\bibnamefont
  {Monticone}},\ }\bibfield  {title} {\bibinfo {title} {Dissipation-induced
  topological transitions in continuous {Weyl} materials},\ }\href
  {https://doi.org/10.1103/PhysRevResearch.2.033065} {\bibfield  {journal}
  {\bibinfo  {journal} {Phys. Rev. Research}\ }\textbf {\bibinfo {volume}
  {2}},\ \bibinfo {pages} {033065} (\bibinfo {year} {2020})}\BibitemShut
  {NoStop}%
\bibitem [{\citenamefont {Yan}\ \emph {et~al.}(2021)\citenamefont {Yan},
  \citenamefont {Chen}, \citenamefont {Zhang}, \citenamefont {Xi},
  \citenamefont {Chen},\ and\ \citenamefont {Yang}}]{Yan2021}%
  \BibitemOpen
  \bibfield  {author} {\bibinfo {author} {\bibfnamefont {Q.}~\bibnamefont
  {Yan}}, \bibinfo {author} {\bibfnamefont {Q.}~\bibnamefont {Chen}}, \bibinfo
  {author} {\bibfnamefont {L.}~\bibnamefont {Zhang}}, \bibinfo {author}
  {\bibfnamefont {R.}~\bibnamefont {Xi}}, \bibinfo {author} {\bibfnamefont
  {H.}~\bibnamefont {Chen}},\ and\ \bibinfo {author} {\bibfnamefont
  {Y.}~\bibnamefont {Yang}},\ }\bibfield  {title} {\bibinfo {title}
  {Unconventional {Weyl} exceptional contours in non-{Hermitian} photonic
  continua},\ }\href {https://doi.org/10.1364/PRJ.438769} {\bibfield  {journal}
  {\bibinfo  {journal} {Photon. Res.}\ }\textbf {\bibinfo {volume} {9}},\
  \bibinfo {pages} {2435} (\bibinfo {year} {2021})}\BibitemShut {NoStop}%
\bibitem [{Sup()}]{SupplementalMaterial}%
  \BibitemOpen
  \href@noop {} {}\bibinfo {note} {See Supplemental Material at [URL will be
  inserted by publisher] for equilibrium details, response-bundle assumptions,
  numerical convergence tests, and relativistic and finite-orbit
  checks.}\BibitemShut {Stop}%
\bibitem [{\citenamefont {Stix}(1992)}]{Stix1992}%
  \BibitemOpen
  \bibfield  {author} {\bibinfo {author} {\bibfnamefont {T.~H.}\ \bibnamefont
  {Stix}},\ }\href@noop {} {\emph {\bibinfo {title} {Waves in Plasmas}}}\
  (\bibinfo  {publisher} {American Institute of Physics},\ \bibinfo {address}
  {New York},\ \bibinfo {year} {1992})\BibitemShut {NoStop}%
\bibitem [{\citenamefont {Verscharen}\ \emph {et~al.}(2018)\citenamefont
  {Verscharen}, \citenamefont {Klein}, \citenamefont {Chandran}, \citenamefont
  {Stevens}, \citenamefont {Salem},\ and\ \citenamefont
  {Bale}}]{Verscharen2018}%
  \BibitemOpen
  \bibfield  {author} {\bibinfo {author} {\bibfnamefont {D.}~\bibnamefont
  {Verscharen}}, \bibinfo {author} {\bibfnamefont {K.~G.}\ \bibnamefont
  {Klein}}, \bibinfo {author} {\bibfnamefont {B.~D.~G.}\ \bibnamefont
  {Chandran}}, \bibinfo {author} {\bibfnamefont {M.~L.}\ \bibnamefont
  {Stevens}}, \bibinfo {author} {\bibfnamefont {C.~S.}\ \bibnamefont {Salem}},\
  and\ \bibinfo {author} {\bibfnamefont {S.~D.}\ \bibnamefont {Bale}},\
  }\bibfield  {title} {\bibinfo {title} {{ALPS}: The arbitrary linear plasma
  solver},\ }\href {https://doi.org/10.1017/S0022377818000739} {\bibfield
  {journal} {\bibinfo  {journal} {J. Plasma Phys.}\ }\textbf {\bibinfo {volume}
  {84}},\ \bibinfo {pages} {905840403} (\bibinfo {year} {2018})}\BibitemShut
  {NoStop}%
\bibitem [{\citenamefont {Fried}\ and\ \citenamefont
  {Conte}(1961)}]{FriedConte1961}%
  \BibitemOpen
  \bibfield  {author} {\bibinfo {author} {\bibfnamefont {B.~D.}\ \bibnamefont
  {Fried}}\ and\ \bibinfo {author} {\bibfnamefont {S.~D.}\ \bibnamefont
  {Conte}},\ }\href {https://doi.org/10.1016/C2013-0-12176-9} {\emph {\bibinfo
  {title} {The Plasma Dispersion Function}}}\ (\bibinfo  {publisher} {Academic
  Press},\ \bibinfo {address} {New York},\ \bibinfo {year} {1961})\BibitemShut
  {NoStop}%
\bibitem [{\citenamefont {Beyn}(2012)}]{Beyn2012}%
  \BibitemOpen
  \bibfield  {author} {\bibinfo {author} {\bibfnamefont {W.-J.}\ \bibnamefont
  {Beyn}},\ }\bibfield  {title} {\bibinfo {title} {An integral method for
  solving nonlinear eigenvalue problems},\ }\href
  {https://doi.org/10.1016/j.laa.2011.03.030} {\bibfield  {journal} {\bibinfo
  {journal} {Linear Algebra Appl.}\ }\textbf {\bibinfo {volume} {436}},\
  \bibinfo {pages} {3839} (\bibinfo {year} {2012})}\BibitemShut {NoStop}%
\bibitem [{\citenamefont {Fukui}\ \emph {et~al.}(2005)\citenamefont {Fukui},
  \citenamefont {Hatsugai},\ and\ \citenamefont {Suzuki}}]{Fukui2005}%
  \BibitemOpen
  \bibfield  {author} {\bibinfo {author} {\bibfnamefont {T.}~\bibnamefont
  {Fukui}}, \bibinfo {author} {\bibfnamefont {Y.}~\bibnamefont {Hatsugai}},\
  and\ \bibinfo {author} {\bibfnamefont {H.}~\bibnamefont {Suzuki}},\
  }\bibfield  {title} {\bibinfo {title} {Chern numbers in discretized brillouin
  zone: Efficient method of computing (spin) hall conductances},\ }\href
  {https://doi.org/10.1143/JPSJ.74.1674} {\bibfield  {journal} {\bibinfo
  {journal} {J. Phys. Soc. Jpn.}\ }\textbf {\bibinfo {volume} {74}},\ \bibinfo
  {pages} {1674} (\bibinfo {year} {2005})}\BibitemShut {NoStop}%
\bibitem [{\citenamefont {Bohlsen}\ \emph {et~al.}(2026)\citenamefont
  {Bohlsen}, \citenamefont {Dodin},\ and\ \citenamefont {Qin}}]{Bohlsen2026}%
  \BibitemOpen
  \bibfield  {author} {\bibinfo {author} {\bibfnamefont {N.}~\bibnamefont
  {Bohlsen}}, \bibinfo {author} {\bibfnamefont {I.~Y.}\ \bibnamefont {Dodin}},\
  and\ \bibinfo {author} {\bibfnamefont {H.}~\bibnamefont {Qin}},\ }\bibfield
  {title} {\bibinfo {title} {Counting topological interface modes using
  simplicial characteristic classes},\ }\href
  {https://doi.org/10.1088/1367-2630/ae384d} {\bibfield  {journal} {\bibinfo
  {journal} {New J. Phys.}\ }\textbf {\bibinfo {volume} {28}},\ \bibinfo
  {pages} {015002} (\bibinfo {year} {2026})}\BibitemShut {NoStop}%
\bibitem [{\citenamefont {Zworski}(2012)}]{Zworski2012}%
  \BibitemOpen
  \bibfield  {author} {\bibinfo {author} {\bibfnamefont {M.}~\bibnamefont
  {Zworski}},\ }\href {https://doi.org/10.1090/gsm/138} {\emph {\bibinfo
  {title} {Semiclassical Analysis}}},\ \bibinfo {series} {Graduate Studies in
  Mathematics}, Vol.\ \bibinfo {volume} {138}\ (\bibinfo  {publisher} {American
  Mathematical Society},\ \bibinfo {address} {Providence, RI},\ \bibinfo {year}
  {2012})\BibitemShut {NoStop}%
\bibitem [{\citenamefont {J{\"u}ttner}(1911)}]{Juttner1911}%
  \BibitemOpen
  \bibfield  {author} {\bibinfo {author} {\bibfnamefont {F.}~\bibnamefont
  {J{\"u}ttner}},\ }\bibfield  {title} {\bibinfo {title} {Das {M}axwellsche
  {G}esetz der {G}eschwindigkeitsverteilung in der {R}elativtheorie},\ }\href
  {https://doi.org/10.1002/andp.19113390503} {\bibfield  {journal} {\bibinfo
  {journal} {Ann. Phys.}\ }\textbf {\bibinfo {volume} {339}},\ \bibinfo {pages}
  {856} (\bibinfo {year} {1911})}\BibitemShut {NoStop}%
\bibitem [{\citenamefont {Yoon}\ and\ \citenamefont
  {Davidson}(1990)}]{YoonDavidson1990}%
  \BibitemOpen
  \bibfield  {author} {\bibinfo {author} {\bibfnamefont {P.~H.}\ \bibnamefont
  {Yoon}}\ and\ \bibinfo {author} {\bibfnamefont {R.~C.}\ \bibnamefont
  {Davidson}},\ }\bibfield  {title} {\bibinfo {title} {Alternative
  representation of the dielectric tensor for a relativistic magnetized plasma
  in thermal equilibrium},\ }\href {https://doi.org/10.1017/S0022377800014781}
  {\bibfield  {journal} {\bibinfo  {journal} {J. Plasma Phys.}\ }\textbf
  {\bibinfo {volume} {43}},\ \bibinfo {pages} {269} (\bibinfo {year}
  {1990})}\BibitemShut {NoStop}%
\bibitem [{\citenamefont {Rajawat}\ \emph {et~al.}(2025)\citenamefont
  {Rajawat}, \citenamefont {Shvets},\ and\ \citenamefont
  {Khudik}}]{Rajawat2025}%
  \BibitemOpen
  \bibfield  {author} {\bibinfo {author} {\bibfnamefont {R.~S.}\ \bibnamefont
  {Rajawat}}, \bibinfo {author} {\bibfnamefont {G.}~\bibnamefont {Shvets}},\
  and\ \bibinfo {author} {\bibfnamefont {V.}~\bibnamefont {Khudik}},\
  }\bibfield  {title} {\bibinfo {title} {Continuum damping of topologically
  protected edge modes at the boundary of cold magnetized plasmas},\ }\href
  {https://doi.org/10.1103/PhysRevLett.134.055301} {\bibfield  {journal}
  {\bibinfo  {journal} {Phys. Rev. Lett.}\ }\textbf {\bibinfo {volume} {134}},\
  \bibinfo {pages} {055301} (\bibinfo {year} {2025})}\BibitemShut {NoStop}%
\bibitem [{\citenamefont {Rao}\ \emph {et~al.}(2026{\natexlab{b}})\citenamefont
  {Rao}, \citenamefont {Li},\ and\ \citenamefont {Sun}}]{RaoZenodo2026}%
  \BibitemOpen
  \bibfield  {author} {\bibinfo {author} {\bibfnamefont {X.}~\bibnamefont
  {Rao}}, \bibinfo {author} {\bibfnamefont {H.}~\bibnamefont {Li}},\ and\
  \bibinfo {author} {\bibfnamefont {X.}~\bibnamefont {Sun}},\ }\href
  {https://doi.org/10.5281/zenodo.22020911} {\bibinfo {title} {{Code and Data
  for Collisionless Resonances Set the Size of a Weyl Exceptional Ring}}}
  (\bibinfo {year} {2026}{\natexlab{b}}),\ \bibinfo {note} {version
  1.0.0}\BibitemShut {NoStop}%
\end{thebibliography}%


\begin{thebibliography}{17}%
\makeatletter
\providecommand \@ifxundefined [1]{%
 \@ifx{#1\undefined}
}%
\providecommand \@ifnum [1]{%
 \ifnum #1\expandafter \@firstoftwo
 \else \expandafter \@secondoftwo
 \fi
}%
\providecommand \@ifx [1]{%
 \ifx #1\expandafter \@firstoftwo
 \else \expandafter \@secondoftwo
 \fi
}%
\providecommand \natexlab [1]{#1}%
\providecommand \enquote  [1]{``#1''}%
\providecommand \bibnamefont  [1]{#1}%
\providecommand \bibfnamefont [1]{#1}%
\providecommand \citenamefont [1]{#1}%
\providecommand \href@noop [0]{\@secondoftwo}%
\providecommand \href [0]{\begingroup \@sanitize@url \@href}%
\providecommand \@href[1]{\@@startlink{#1}\@@href}%
\providecommand \@@href[1]{\endgroup#1\@@endlink}%
\providecommand \@sanitize@url [0]{\catcode `\\12\catcode `\$12\catcode
  `\&12\catcode `\#12\catcode `\^12\catcode `\_12\catcode `\%12\relax}%
\providecommand \@@startlink[1]{}%
\providecommand \@@endlink[0]{}%
\providecommand \url  [0]{\begingroup\@sanitize@url \@url }%
\providecommand \@url [1]{\endgroup\@href {#1}{\urlprefix }}%
\providecommand \urlprefix  [0]{URL }%
\providecommand \Eprint [0]{\href }%
\providecommand \doibase [0]{https://doi.org/}%
\providecommand \selectlanguage [0]{\@gobble}%
\providecommand \bibinfo  [0]{\@secondoftwo}%
\providecommand \bibfield  [0]{\@secondoftwo}%
\providecommand \translation [1]{[#1]}%
\providecommand \BibitemOpen [0]{}%
\providecommand \bibitemStop [0]{}%
\providecommand \bibitemNoStop [0]{.\EOS\space}%
\providecommand \EOS [0]{\spacefactor3000\relax}%
\providecommand \BibitemShut  [1]{\csname bibitem#1\endcsname}%
\let\auto@bib@innerbib\@empty
\bibitem [{\citenamefont {Verscharen}\ \emph {et~al.}(2018)\citenamefont
  {Verscharen}, \citenamefont {Klein}, \citenamefont {Chandran}, \citenamefont
  {Stevens}, \citenamefont {Salem},\ and\ \citenamefont
  {Bale}}]{Verscharen2018}%
  \BibitemOpen
  \bibfield  {author} {\bibinfo {author} {\bibfnamefont {D.}~\bibnamefont
  {Verscharen}}, \bibinfo {author} {\bibfnamefont {K.~G.}\ \bibnamefont
  {Klein}}, \bibinfo {author} {\bibfnamefont {B.~D.~G.}\ \bibnamefont
  {Chandran}}, \bibinfo {author} {\bibfnamefont {M.~L.}\ \bibnamefont
  {Stevens}}, \bibinfo {author} {\bibfnamefont {C.~S.}\ \bibnamefont {Salem}},\
  and\ \bibinfo {author} {\bibfnamefont {S.~D.}\ \bibnamefont {Bale}},\
  }\bibfield  {title} {\bibinfo {title} {{ALPS}: The arbitrary linear plasma
  solver},\ }\href {https://doi.org/10.1017/S0022377818000739} {\bibfield
  {journal} {\bibinfo  {journal} {J. Plasma Phys.}\ }\textbf {\bibinfo {volume}
  {84}},\ \bibinfo {pages} {905840403} (\bibinfo {year} {2018})}\BibitemShut
  {NoStop}%
\bibitem [{\citenamefont {Fukui}\ \emph {et~al.}(2005)\citenamefont {Fukui},
  \citenamefont {Hatsugai},\ and\ \citenamefont {Suzuki}}]{Fukui2005}%
  \BibitemOpen
  \bibfield  {author} {\bibinfo {author} {\bibfnamefont {T.}~\bibnamefont
  {Fukui}}, \bibinfo {author} {\bibfnamefont {Y.}~\bibnamefont {Hatsugai}},\
  and\ \bibinfo {author} {\bibfnamefont {H.}~\bibnamefont {Suzuki}},\
  }\bibfield  {title} {\bibinfo {title} {Chern numbers in discretized brillouin
  zone: Efficient method of computing (spin) hall conductances},\ }\href
  {https://doi.org/10.1143/JPSJ.74.1674} {\bibfield  {journal} {\bibinfo
  {journal} {J. Phys. Soc. Jpn.}\ }\textbf {\bibinfo {volume} {74}},\ \bibinfo
  {pages} {1674} (\bibinfo {year} {2005})}\BibitemShut {NoStop}%
\bibitem [{\citenamefont {Bohlsen}\ \emph {et~al.}(2026)\citenamefont
  {Bohlsen}, \citenamefont {Dodin},\ and\ \citenamefont {Qin}}]{Bohlsen2026}%
  \BibitemOpen
  \bibfield  {author} {\bibinfo {author} {\bibfnamefont {N.}~\bibnamefont
  {Bohlsen}}, \bibinfo {author} {\bibfnamefont {I.~Y.}\ \bibnamefont {Dodin}},\
  and\ \bibinfo {author} {\bibfnamefont {H.}~\bibnamefont {Qin}},\ }\bibfield
  {title} {\bibinfo {title} {Counting topological interface modes using
  simplicial characteristic classes},\ }\href
  {https://doi.org/10.1088/1367-2630/ae384d} {\bibfield  {journal} {\bibinfo
  {journal} {New J. Phys.}\ }\textbf {\bibinfo {volume} {28}},\ \bibinfo
  {pages} {015002} (\bibinfo {year} {2026})}\BibitemShut {NoStop}%
\bibitem [{\citenamefont {Raman}\ and\ \citenamefont
  {Fan}(2010)}]{RamanFan2010}%
  \BibitemOpen
  \bibfield  {author} {\bibinfo {author} {\bibfnamefont {A.}~\bibnamefont
  {Raman}}\ and\ \bibinfo {author} {\bibfnamefont {S.}~\bibnamefont {Fan}},\
  }\bibfield  {title} {\bibinfo {title} {Photonic band structure of dispersive
  metamaterials formulated as a hermitian eigenvalue problem},\ }\href
  {https://doi.org/10.1103/PhysRevLett.104.087401} {\bibfield  {journal}
  {\bibinfo  {journal} {Phys. Rev. Lett.}\ }\textbf {\bibinfo {volume} {104}},\
  \bibinfo {pages} {087401} (\bibinfo {year} {2010})}\BibitemShut {NoStop}%
\bibitem [{\citenamefont {Silveirinha}(2015)}]{Silveirinha2015}%
  \BibitemOpen
  \bibfield  {author} {\bibinfo {author} {\bibfnamefont {M.~G.}\ \bibnamefont
  {Silveirinha}},\ }\bibfield  {title} {\bibinfo {title} {Chern invariants for
  continuous media},\ }\href {https://doi.org/10.1103/PhysRevB.92.125153}
  {\bibfield  {journal} {\bibinfo  {journal} {Phys. Rev. B}\ }\textbf {\bibinfo
  {volume} {92}},\ \bibinfo {pages} {125153} (\bibinfo {year}
  {2015})}\BibitemShut {NoStop}%
\bibitem [{\citenamefont {Chiba}(2015)}]{Chiba2015}%
  \BibitemOpen
  \bibfield  {author} {\bibinfo {author} {\bibfnamefont {H.}~\bibnamefont
  {Chiba}},\ }\bibfield  {title} {\bibinfo {title} {A spectral theory of linear
  operators on rigged {Hilbert} spaces under analyticity conditions},\ }\href
  {https://doi.org/10.1016/j.aim.2015.01.001} {\bibfield  {journal} {\bibinfo
  {journal} {Adv. Math.}\ }\textbf {\bibinfo {volume} {273}},\ \bibinfo {pages}
  {324} (\bibinfo {year} {2015})}\BibitemShut {NoStop}%
\bibitem [{\citenamefont {Prigogine}\ and\ \citenamefont
  {Petrosky}(1998)}]{PrigoginePetrosky1998}%
  \BibitemOpen
  \bibfield  {author} {\bibinfo {author} {\bibfnamefont {I.}~\bibnamefont
  {Prigogine}}\ and\ \bibinfo {author} {\bibfnamefont {T.}~\bibnamefont
  {Petrosky}},\ }\bibfield  {title} {\bibinfo {title} {Semigroup representation
  of the {Vlasov} evolution},\ }\href
  {https://doi.org/10.1017/S002237789800662X} {\bibfield  {journal} {\bibinfo
  {journal} {J. Plasma Phys.}\ }\textbf {\bibinfo {volume} {59}},\ \bibinfo
  {pages} {611} (\bibinfo {year} {1998})}\BibitemShut {NoStop}%
\bibitem [{\citenamefont {Fried}\ and\ \citenamefont
  {Conte}(1961)}]{FriedConte1961}%
  \BibitemOpen
  \bibfield  {author} {\bibinfo {author} {\bibfnamefont {B.~D.}\ \bibnamefont
  {Fried}}\ and\ \bibinfo {author} {\bibfnamefont {S.~D.}\ \bibnamefont
  {Conte}},\ }\href {https://doi.org/10.1016/C2013-0-12176-9} {\emph {\bibinfo
  {title} {The Plasma Dispersion Function}}}\ (\bibinfo  {publisher} {Academic
  Press},\ \bibinfo {address} {New York},\ \bibinfo {year} {1961})\BibitemShut
  {NoStop}%
\bibitem [{\citenamefont {van Kampen}(1955)}]{VanKampen1955}%
  \BibitemOpen
  \bibfield  {author} {\bibinfo {author} {\bibfnamefont {N.~G.}\ \bibnamefont
  {van Kampen}},\ }\bibfield  {title} {\bibinfo {title} {On the theory of
  stationary waves in plasmas},\ }\href
  {https://doi.org/10.1016/S0031-8914(55)93068-8} {\bibfield  {journal}
  {\bibinfo  {journal} {Physica}\ }\textbf {\bibinfo {volume} {21}},\ \bibinfo
  {pages} {949} (\bibinfo {year} {1955})}\BibitemShut {NoStop}%
\bibitem [{\citenamefont {Case}(1959)}]{Case1959}%
  \BibitemOpen
  \bibfield  {author} {\bibinfo {author} {\bibfnamefont {K.~M.}\ \bibnamefont
  {Case}},\ }\bibfield  {title} {\bibinfo {title} {Plasma oscillations},\
  }\href {https://doi.org/10.1016/0003-4916(59)90029-6} {\bibfield  {journal}
  {\bibinfo  {journal} {Ann. Phys.}\ }\textbf {\bibinfo {volume} {7}},\
  \bibinfo {pages} {349} (\bibinfo {year} {1959})}\BibitemShut {NoStop}%
\bibitem [{\citenamefont {Ramos}(2019)}]{Ramos2019}%
  \BibitemOpen
  \bibfield  {author} {\bibinfo {author} {\bibfnamefont {J.~J.}\ \bibnamefont
  {Ramos}},\ }\bibfield  {title} {\bibinfo {title} {Normal-mode-based theory of
  collisionless plasma waves},\ }\href
  {https://doi.org/10.1017/S0022377819000400} {\bibfield  {journal} {\bibinfo
  {journal} {J. Plasma Phys.}\ }\textbf {\bibinfo {volume} {85}},\ \bibinfo
  {pages} {905850401} (\bibinfo {year} {2019})}\BibitemShut {NoStop}%
\bibitem [{\citenamefont {Beyn}(2012)}]{Beyn2012}%
  \BibitemOpen
  \bibfield  {author} {\bibinfo {author} {\bibfnamefont {W.-J.}\ \bibnamefont
  {Beyn}},\ }\bibfield  {title} {\bibinfo {title} {An integral method for
  solving nonlinear eigenvalue problems},\ }\href
  {https://doi.org/10.1016/j.laa.2011.03.030} {\bibfield  {journal} {\bibinfo
  {journal} {Linear Algebra Appl.}\ }\textbf {\bibinfo {volume} {436}},\
  \bibinfo {pages} {3839} (\bibinfo {year} {2012})}\BibitemShut {NoStop}%
\bibitem [{\citenamefont {Faure}(2023)}]{Faure2023}%
  \BibitemOpen
  \bibfield  {author} {\bibinfo {author} {\bibfnamefont {F.}~\bibnamefont
  {Faure}},\ }\bibfield  {title} {\bibinfo {title} {Manifestation of the
  topological index formula in quantum waves and geophysical waves},\ }\href
  {https://doi.org/10.5802/ahl.169} {\bibfield  {journal} {\bibinfo  {journal}
  {Ann. Henri Lebesgue}\ }\textbf {\bibinfo {volume} {6}},\ \bibinfo {pages}
  {449} (\bibinfo {year} {2023})}\BibitemShut {NoStop}%
\bibitem [{\citenamefont {Qin}\ and\ \citenamefont {Fu}(2023)}]{QinFu2023}%
  \BibitemOpen
  \bibfield  {author} {\bibinfo {author} {\bibfnamefont {H.}~\bibnamefont
  {Qin}}\ and\ \bibinfo {author} {\bibfnamefont {Y.}~\bibnamefont {Fu}},\
  }\bibfield  {title} {\bibinfo {title} {Topological langmuir-cyclotron wave},\
  }\href {https://doi.org/10.1126/sciadv.add8041} {\bibfield  {journal}
  {\bibinfo  {journal} {Sci. Adv.}\ }\textbf {\bibinfo {volume} {9}},\ \bibinfo
  {pages} {eadd8041} (\bibinfo {year} {2023})}\BibitemShut {NoStop}%
\bibitem [{\citenamefont {J{\"u}ttner}(1911)}]{Juttner1911}%
  \BibitemOpen
  \bibfield  {author} {\bibinfo {author} {\bibfnamefont {F.}~\bibnamefont
  {J{\"u}ttner}},\ }\bibfield  {title} {\bibinfo {title} {Das {M}axwellsche
  {G}esetz der {G}eschwindigkeitsverteilung in der {R}elativtheorie},\ }\href
  {https://doi.org/10.1002/andp.19113390503} {\bibfield  {journal} {\bibinfo
  {journal} {Ann. Phys.}\ }\textbf {\bibinfo {volume} {339}},\ \bibinfo {pages}
  {856} (\bibinfo {year} {1911})}\BibitemShut {NoStop}%
\bibitem [{\citenamefont {Yoon}\ and\ \citenamefont
  {Davidson}(1990)}]{YoonDavidson1990}%
  \BibitemOpen
  \bibfield  {author} {\bibinfo {author} {\bibfnamefont {P.~H.}\ \bibnamefont
  {Yoon}}\ and\ \bibinfo {author} {\bibfnamefont {R.~C.}\ \bibnamefont
  {Davidson}},\ }\bibfield  {title} {\bibinfo {title} {Alternative
  representation of the dielectric tensor for a relativistic magnetized plasma
  in thermal equilibrium},\ }\href {https://doi.org/10.1017/S0022377800014781}
  {\bibfield  {journal} {\bibinfo  {journal} {J. Plasma Phys.}\ }\textbf
  {\bibinfo {volume} {43}},\ \bibinfo {pages} {269} (\bibinfo {year}
  {1990})}\BibitemShut {NoStop}%
\bibitem [{\citenamefont {Morrison}(1980)}]{Morrison1980}%
  \BibitemOpen
  \bibfield  {author} {\bibinfo {author} {\bibfnamefont {P.~J.}\ \bibnamefont
  {Morrison}},\ }\bibfield  {title} {\bibinfo {title} {The {Maxwell--Vlasov}
  equations as a continuous {Hamiltonian} system},\ }\href
  {https://doi.org/10.1016/0375-9601(80)90776-8} {\bibfield  {journal}
  {\bibinfo  {journal} {Phys. Lett. A}\ }\textbf {\bibinfo {volume} {80}},\
  \bibinfo {pages} {383} (\bibinfo {year} {1980})}\BibitemShut {NoStop}%
\end{thebibliography}%

\end{document}


\renewcommand{\thefigure}{S\arabic{figure}}
\title{Supplemental Material for ``Collisionless Resonances Set the Size of a Weyl Exceptional Ring''}
\author{Xianhao Rao}
\email{rrxxhh@mail.ustc.edu.cn}
\affiliation{Xeonova Ltd., Hefei 230093, People's Republic of China}

\author{Hong Li}
\email{honglee@ustc.edu.cn}
\affiliation{Xeonova Ltd., Hefei 230093, People's Republic of China}
\affiliation{School of Nuclear Science and Technology, University of Science and Technology of China, No. 443 Huangshan Road, Hefei, Anhui, People's Republic of China}

\author{Xuan Sun}
\email{xsun@ustc.edu.cn}
\affiliation{Xeonova Ltd., Hefei 230093, People's Republic of China}
\affiliation{School of Nuclear Science and Technology, University of Science and Technology of China, No. 443 Huangshan Road, Hefei, Anhui, People's Republic of China}
\date{}
\maketitle

\section{Exact equilibrium and principal-symbol reduction}

The baseline calculation uses the nonrelativistic Vlasov--Maxwell system.  The index $s$ labels a physical species, and $a\in\{1,2\}$ labels its Maxwellian components.  For charge $q_s$, mass $m_s$, and temperature $T_{sa}$, define
\begin{equation}
 \Omega_s=\frac{q_sB_0}{m_s},\qquad
 v_{tsa}^2=\frac{2T_{sa}}{m_s},\qquad
 \rho_{sa}^{2}=\frac{T_{sa}}{m_s\Omega_s^2}
 =\frac{\langle v_y^2\rangle_{sa}}{\Omega_s^2}.
\end{equation}
Here $\Omega_s$ is the signed cyclotron frequency and $\rho_{sa}$ is the
one-coordinate rms orbit-displacement width.  With the thermal-speed
convention above, the commonly used thermal Larmor radius is
$v_{tsa}/|\Omega_s|=\sqrt2\rho_{sa}$.  In a uniform magnetic field,
\begin{equation}
 \mathcal X_s=x+v_y/\Omega_s
\end{equation}
is an invariant of the unperturbed characteristics, as is $v^2$.  Hence
\begin{equation}
 f_{0s}(x,\bm v)=\sum_{a=1}^{2}N_{sa}(\mathcal X_s)
 (\pi v_{tsa}^2)^{-3/2}e^{-v^2/v_{tsa}^2}
 \label{S:eq}
\end{equation}
solves the stationary collisionless Vlasov equation exactly.

We prescribe the physical total density
\begin{equation}
 n(x)=n_-+(n_+-n_-)\frac{1+\operatorname{erf}(x/L)}2
\end{equation}
and split it into two positive Maxwellian components at constant pressure $\bar p_s$:
\begin{equation}
 n_{s1}(x)=\frac{T_{s2}n(x)-\bar p_s}{T_{s2}-T_{s1}},\qquad
 n_{s2}(x)=\frac{\bar p_s-T_{s1}n(x)}{T_{s2}-T_{s1}}.
 \label{S:split}
\end{equation}
Then $n_{s1}+n_{s2}=n$ and $T_{s1}n_{s1}+T_{s2}n_{s2}=\bar p_s$.  Positivity requires
\begin{equation}
 T_{s1}\max(n_-,n_+)<\bar p_s<T_{s2}\min(n_-,n_+).
\end{equation}
We take $T_{s2}/T_{s1}=10$ and $\bar p_s=2T_{s1}\max(n_-,n_+)$.  The
baseline ion completion uses $(T_{i1},T_{i2})=(0.08,0.8)\,$keV.

Velocity integration of $N_{sa}(x+v_y/\Omega_s)$ is a Gaussian convolution of variance $\rho_{sa}^2$.  If a term in Eq.~(\ref{S:split}) is $A_{sa}+B_{sa}\operatorname{erf}(x/L)$, its exact preconvolution guiding-center profile is
\begin{equation}
 N_{sa}(\mathcal X)=A_{sa}+B_{sa}\operatorname{erf}
 \left(\frac{\mathcal X}{\sqrt{L^2-2\rho_{sa}^2}}\right).
 \label{S:precursor}
\end{equation}
Equation~(\ref{S:eq}) generates the prescribed physical density provided $L^2>2\rho_{sa}^2$.  For the parameters used in the Letter, the electron rms orbit widths are $6.74\times10^{-5}\,$m and $2.13\times10^{-4}\,$m; the proton values are $9.14\times10^{-4}\,$m and $2.89\times10^{-3}\,$m.  All positivity and width inequalities are strict.

\subsection{Maxwell self-consistency of the exact equilibrium}

Maxwell self-consistency also requires the moments of Eq.~(\ref{S:eq}) to support the prescribed fields.  Gaussian integration gives $n_{sa}(x)=\int f_{0sa}\,d^3v$ and, after integration by parts in $v_y$,
\begin{align}
 \Phi_{y,sa}
 &=\int v_y f_{0sa}\,d^3v
 =\frac{T_{sa}}{m_s\Omega_s}\frac{dn_{sa}}{dx},\\
 j_{y,sa}&=q_s\Phi_{y,sa}=\frac{T_{sa}}{B_0}\frac{dn_{sa}}{dx}.
 \label{S:diamagnetic-current}
\end{align}
The two components were chosen so that $\sum_aT_{sa}n_{sa}=\bar p_s$ for each species.  Consequently
\begin{equation}
 j_{0y}=\frac{1}{B_0}\frac{d}{dx}\sum_{s,a}T_{sa}n_{sa}=0,
 \qquad j_{0x}=j_{0z}=0.
 \label{S:current-cancel}
\end{equation}
Taking the electron and proton physical densities equal also gives $\rho_0=\sum_sq_sn_s=0$.  Hence $\bm E_0=0$ and uniform $\bm B_0$ satisfy both $\nabla\cdot\bm E_0=\rho_0/\epsilon_0$ and $\nabla\times\bm B_0=\mu_0\bm j_0$.  The individual component diamagnetic currents are finite, but pressure balance cancels their sum exactly.

\subsection{From the exact equilibrium to the frozen principal symbol}

Let $G_{\rho}$ denote convolution with the one-dimensional Gaussian of variance $\rho^2$.  The velocity integral in Eq.~(\ref{S:eq}) is exactly
\begin{equation}
 n_{sa}=G_{\rho_{sa}}N_{sa}
 =\exp\!\left(\frac{\rho_{sa}^2}{2}\partial_x^2\right)N_{sa},
 \qquad
 N_{sa}=n_{sa}-\frac{\rho_{sa}^2}{2}n_{sa}''
 +O(\rho_{sa}^4\partial_x^4n_{sa}).
 \label{S:heat-kernel}
\end{equation}
Expanding the exact phase-space profile at fixed $x$ then yields
\begin{equation}
 N_{sa}\!\left(x+\frac{v_y}{\Omega_s}\right)
 =n_{sa}(x)+\frac{v_y}{\Omega_s}n_{sa}'(x)
 +\frac12\!\left[\left(\frac{v_y}{\Omega_s}\right)^2-\rho_{sa}^2\right]n_{sa}''(x)
 +O\!\left[n_{sa}(\rho_{sa}/L)^3\right].
 \label{S:equilibrium-expansion}
\end{equation}
The remainder in Eq.~(\ref{S:equilibrium-expansion}) is taken in the Maxwellian-weighted moment expansion.  The first term is the frozen isotropic Maxwellian used in the susceptibility below.  The odd $O(\rho_{sa}/L)$ term carries the component diamagnetic drift in Eq.~(\ref{S:diamagnetic-current}), while the even amplitude correction starts at $O[(\rho_{sa}/L)^2]$.  We use $X=x/L$, $w=z/\omega_{ce}$, and $\bm u=c\bm k/\omega_{ce}$, where $z$ is the complex angular frequency and $\omega_{ce}=eB_0/m_e>0$.  Freezing the first term in a Wigner expansion and replacing $k_x$ by the phase-space covariable gives the retarded electric-field principal symbol $\Dret(w,\bm u,X)$.  Orbit sampling of the odd term and Moyal brackets generated by the $X$ dependence enter at subprincipal order.  Since
\begin{equation}
 \frac{\rho_{sa}}{L}=\frac{v_{tsa}}{\sqrt2c}\,
 \frac{c}{|\Omega_s|L},
\end{equation}
magnetization and the semiclassical scale control this hierarchy independently.  The frozen tensor is the leading symbol of the exact equilibrium; orbit sampling and gradient corrections can shift the defect and pole at the next order.

\section{Retarded hot-plasma susceptibility}

We use fields proportional to $e^{\ii\bm k\cdot\bm x-\ii zt}$ and write $k_\parallel=k_z$, $v_\parallel=v_z$, $k_\perp=(k_x^2+k_y^2)^{1/2}$, and $v_\perp=(v_x^2+v_y^2)^{1/2}$.  A superscript $+$ denotes retarded analytic continuation.  For Maxwellian component $(s,a)$, define the cyclotron-harmonic argument
\begin{equation}
 Z^+(\zeta)=\ii\sqrt\pi\,w_{\rm F}(\zeta),\qquad
 \zeta_{\ell,sa}=\frac{z-\ell\Omega_s}{k_\parallel v_{tsa}},
 \qquad \ell\in\mathbb Z,
\end{equation}
where $w_{\rm F}$ is the entire Faddeeva function.  This expression is the retarded Landau-contour continuation of the velocity integral.  Direct evaluation of the original real-axis integral at $\operatorname{Im}z<0$ would not reproduce that continuation.

For a single isotropic Maxwellian component, the susceptibility used in the code is the nonrelativistic specialization of the general Vlasov tensor in Ref.~\cite{Verscharen2018}.  For $k_\parallel\ne0$, as in the calculations reported here, set
\begin{equation}
 \eta=\frac{v_\perp^2}{v_{tsa}^2},\qquad
 \lambda=\frac{k_\perp v_{tsa}\sqrt\eta}{\Omega_s},\qquad
 J=J_\ell(\lambda),\qquad J'=\partial_\lambda J_\ell(\lambda),
 \qquad b=\frac{\ell J}{\lambda},
\end{equation}
where $J_\ell$ is the Bessel function of the first kind.  With $\zeta=\zeta_{\ell,sa}$, define
\begin{align}
 I_0&=-\frac{Z^+(\zeta)}{k_\parallel v_{tsa}},&
 I_1&=-\frac{1+\zeta Z^+(\zeta)}{k_\parallel v_{tsa}},\\
 I_2&=-\frac{\zeta[1+\zeta Z^+(\zeta)]}{k_\parallel v_{tsa}}.
\end{align}
The component plasma frequency is $\omega_{p,sa}^2=n_{sa}q_s^2/(\epsilon_0m_s)$.  Its susceptibility is
\begin{equation}
 \bm\chi_{sa}^+=-\frac{2\omega_{p,sa}^2}{z}
 \sum_{\ell=-\infty}^{\infty}\int_0^\infty e^{-\eta}\,
 \mathsf T_{\ell}(\eta)\,d\eta,
 \label{S:chi}
\end{equation}
where
\begin{equation}
\mathsf T_{\ell}=
\begin{pmatrix}
\eta b^2I_0&\ii\eta bJ'I_0&\sqrt{\eta}bJI_1\\
-\ii\eta bJ'I_0&\eta J'^2I_0&-\ii\sqrt{\eta}JJ'I_1\\
\sqrt{\eta}bJI_1&\ii\sqrt{\eta}JJ'I_1&J^2I_2
\end{pmatrix}.
\end{equation}
The total dielectric tensor is $\bm\varepsilon^+=\bm1+\sum_{s,a}\bm\chi_{sa}^+$ over the species included in the response.  The $\eta$ integral is evaluated by generalized Gauss--Laguerre quadrature.  For each component, let
\begin{equation}
 \Lambda_{sa}=\max_j\left|
 \frac{k_\perp v_{tsa}\sqrt{\eta_j}}{\Omega_s}\right|,
 \qquad
 \ell_{\max,sa}=\left\lceil
 \Lambda_{sa}+4\sqrt{\Lambda_{sa}+1}\right\rceil+8,
\end{equation}
where $\eta_j$ are the quadrature nodes.  Since $\Lambda_{sa}\geq0$, the outer $\max(3,\cdot)$ retained in the code is redundant for the displayed padding of 8.  This conservative cutoff is an empirical safety rule rather than an analytic tail bound.  We check it by increasing the quadrature order and harmonic padding independently.  At the baseline point, padding values 4, 8, and 12 give the same exceptional root to machine precision.
Removable Bessel ratios on the symmetry axis are taken by their analytic limits.  At quadrature order 16, a $10^{-5}\,$eV test at a generic complex frequency agrees with the Cartesian cold susceptibility to $1.8\times10^{-6}$.

$Z^+$ implements the physical retarded continuation of the Vlasov response.  Gauss--Laguerre and Fourier cutoffs are refined as numerical convergence parameters.  We introduce no collision shift $z\to z+\ii\nu$, Pad\'e approximation to $Z$, or deletion of resonant denominators.

The same formula explains physically why the cold crossing cannot simply acquire one common linewidth.  With the Fourier convention stated above, define $\bm e_\pm=(\hat{\bm x}\pm\ii\hat{\bm y})/\sqrt2$.  For $\bm B_0=B_0\hat{\bm z}$, the $\bm e_+$ sector is the right-hand circularly polarized ($R$), electron-cyclotron-resonant branch.  At $k_\perp=0$, the longitudinal and $R$-polarized sectors decouple, but they project onto different combinations of the resonant denominators
\begin{equation}
 z-k_\parallel v_\parallel-\ell\Omega_s=0.
\end{equation}
The Langmuir-like root is weighted mainly by the parallel response, whereas the $R$-polarized root samples the cyclotron-shifted harmonics.  Their imaginary self-energies therefore need not agree even when their real frequencies are tuned into coincidence.  In a local two-root description this differential damping appears as an imaginary component of $d_z$.  The off-axis coupling supplies $d_x$ and $d_y$, allowing $\bm d\cdot\bm d=0$ at finite $k_\perp$.  Thus the ring radius measures the competition between polarization mixing and unequal collisionless damping; it is not a numerical broadening scale.

\section{Differential damping and the exceptional-ring radius}

The symmetry axis gives an independent way to predict the ring before solving
the off-axis exceptional-root problem.  Write
\begin{equation}
 K=\frac{ck_z}{\omega_{ce}},\qquad
 \beta_a=\frac{v_{ta}}{c},\qquad
 \nu_a=\frac{\omega_{pa}^2}{\omega_{ce}^2}.
\end{equation}
For the two electron Maxwellian components, the $R$-polarized and longitudinal
scalar equations obtained from the full tensor at $k_\perp=0$ are
\begin{align}
 m_+(w)&=1-\frac{K^2}{w^2}
 +\sum_a\frac{\nu_a}{wK\beta_a}
 Z^+\!\left(\frac{w-1}{K\beta_a}\right),
 \label{S:axis-plus}\\
 m_\parallel(w)&=1+\sum_a\frac{2\nu_a}{wK\beta_a}
 \zeta_{a0}\left[1+\zeta_{a0}Z^+(\zeta_{a0})\right],
 \qquad \zeta_{a0}=\frac{w}{K\beta_a}.
 \label{S:axis-parallel}
\end{align}
Equation~(\ref{S:axis-plus}) contains the electron harmonic compatible with
the selected $R$ polarization, while Eq.~(\ref{S:axis-parallel})
contains the $\ell=0$ Landau response.  We also evaluate $Z^+$ directly as
\begin{equation}
 Z^+(\zeta)=\frac{1}{\sqrt\pi}
 \int_{\mathcal C_L}\frac{e^{-q^2}}{q-\zeta}\,dq,
 \label{S:landau-contour}
\end{equation}
where the contour $\mathcal C_L$ passes below the continuously followed pole.
For the baseline roots, this contour agrees with the Faddeeva representation to better than $6\times10^{-16}$ and verifies the causal sheet.

Let $X_0$ be chosen so that
$\operatorname{Re}w_+(X_0)=\operatorname{Re}w_\parallel(X_0)$.  With
$m_j=m_j^{(0)}+\sum_a\chi_j^{(a)}$ and
$\Gamma_j=-\operatorname{Im}w_j$ (so the dimensional damping rate is
$\omega_{ce}\Gamma_j$), weak-damping perturbation theory gives the
component-resolved functional
\begin{equation}
 \Gamma_j^{(a)}=
 \frac{\operatorname{Im}\chi_j^{(a)}(w_r)}
 {\partial_w\operatorname{Re}m_j(w_r)},
 \qquad
 \Delta\Gamma=\sum_a\left(\Gamma_+^{(a)}-
 \Gamma_\parallel^{(a)}\right).
 \label{S:differential-damping}
\end{equation}
On the real axis, $\operatorname{Im}Z^+(\zeta)=\sqrt\pi e^{-\zeta^2}$.
Equation~(\ref{S:differential-damping}) therefore combines the resonant-particle population with the wave-energy normalization supplied by the root derivative.  The distribution value at one resonant velocity captures only the first factor.

The exact Schur complement introduced below is diagonal on the axis.  Expanding
each diagonal entry about its own complex root and keeping the leading
off-axis coupling gives
\begin{equation}
 \mathsf M^+\simeq
 \begin{pmatrix}
 m_+'(w_+)(w-w_+)&u_\perp g_{+z}\\
 u_\perp g_{z+}&m_\parallel'(w_\parallel)(w-w_\parallel)
 \end{pmatrix}.
 \label{S:root-centered-matrix}
\end{equation}
The corresponding discriminant is
\begin{equation}
 (w_+-w_\parallel)^2+4u_\perp^2v_{\rm mix}^2=0,
 \qquad
 v_{\rm mix}^2=\frac{g_{+z}g_{z+}}
 {m_+'(w_+)m_\parallel'(w_\parallel)}.
 \label{S:root-centered-discriminant}
\end{equation}
The product in Eq.~(\ref{S:root-centered-discriminant}) is complex bilinear;
it is not an absolute square.  When the real parts of the two roots coincide,
the leading radius is
\begin{equation}
 R_{\rm EP}^{\rm pred}=\frac{|\Delta\Gamma|}{2|v_{\rm mix}|}.
 \label{S:radius-scaling}
\end{equation}
For a complex coupling phase, a real shift $\delta X$ is determined together
with $R_{\rm lin}$ from
\begin{equation}
 [\Delta w+\partial_X(\Delta w)\delta X]^2
 +4v_{\rm mix}^2R_{\rm lin}^2=0.
 \label{S:complex-linear-scaling}
\end{equation}
Here $\Delta w=w_+-w_\parallel$, and the quantities without the shifted
argument are evaluated at $X_0$.
This is the complex root-centered prediction used for the open symbols in
Fig.~1(c) of the Letter.

For the baseline parameters, the independently solved axial roots and mixing
coefficient are
\begin{align}
 w_+&=0.617784-1.5510\times10^{-3}\ii,\\
 w_\parallel&=0.617784-6.290\times10^{-5}\ii,\\
 v_{\rm mix}&=0.3011-0.0054\ii.
\end{align}
They give $\Delta\Gamma=1.4881\times10^{-3}$,
$R_{\rm EP}^{\rm pred}=2.4705\times10^{-3}$, and the full nonlinear conditions give
$R_{\rm EP}^{\rm full}=2.4701\times10^{-3}$.  The 8-keV component contributes more
than $99.9\%$ of the $R$-branch damping.  Its cyclotron and Landau resonance
arguments are $-2.16$ and $3.49$; for the 0.8-keV component they are $-6.83$
and $11.0$.

The scaling archive contains 26 unique parameter points: $T_{e1}=0.4$--$1.2$
keV at fixed ratio 10, $K=0.85$--$1.15$, $B_0=0.85$--$1.15$ T, a common
density/pressure factor 0.8--1.2, and temperature ratios 8--15.  Across this
set, $|\Delta\Gamma|$ spans a factor of 310 and $R_{\rm EP}^{\rm full}$ a factor of
343.  The maximum relative error of Eq.~(\ref{S:radius-scaling}) is
$1.17\times10^{-3}$; Eq.~(\ref{S:complex-linear-scaling}) reduces it to
$2.68\times10^{-4}$.  Replacing $\Delta\Gamma$ by the mean damping gives only
50--94\% of the full radius and does not collapse the data.

For a scanned parameter set $\bm\theta_{\rm par}$, write
\begin{equation}
 R_{\rm EP}^{\rm full}=R_{\rm lin}+\alpha_2(\bm\theta_{\rm par})R_{\rm lin}^2
 +O(R_{\rm lin}^3),
\end{equation}
the scan gives $-0.045\lesssim\alpha_2\lesssim0.020$.  Its sign change reflects parameter-dependent cone curvature, frequency dispersion, and variation of the mixing matrix.  All 26 full roots have geometric nullity one.  The largest double-root residual is $9.1\times10^{-11}$, and increasing the Gauss--Laguerre order from 12 to 32 changes the baseline radius by less than $7\times10^{-12}$ relative.

\subsection{Nonlinear exceptional-ring verification}

Let $U=(\bm e_+,\bm e_z,\bm e_-)$ with $\bm e_\pm=(1,\pm\ii,0)^T/\sqrt2$.  In this basis write
\begin{equation}
 U^\dagger\Dret U=
 \begin{pmatrix}\mathsf A&\mathsf B\\\mathsf C&\mathsf S\end{pmatrix},
 \qquad
 \mathsf M^+=\mathsf A-\mathsf B\mathsf S^{-1}\mathsf C.
\end{equation}
Here $\mathsf A$ is the $2\times2$ block associated with $(\bm e_+,\bm e_z)$ and $\mathsf S$ is the scalar $\bm e_-$ block.  On all surfaces used below $|\mathsf S|>1.98$, so this is an exact elimination of the separated polarization.  At the reference point $p_*=(w_*,0,0,X_*)$, define
\begin{equation}
 M_0=\mathsf M^+(p_*),\quad
 M_w=\left.\partial_w\mathsf M^+\right|_{p_*},\quad
 M_x=\left.\partial_{u_x}\mathsf M^+\right|_{p_*},\quad
 M_y=\left.\partial_{u_y}\mathsf M^+\right|_{p_*},\quad
 M_X=\left.\partial_X\mathsf M^+\right|_{p_*}.
\end{equation}
The local frequency-linearized Hamiltonian is
\begin{equation}
 H=w_*\bm1-M_w^{-1}\left[M_0+M_xu_x+M_yu_y+M_X(X-X_*)\right].
\end{equation}
Equivalently, $H=H_0+H_xu_x+H_yu_y+H_X(X-X_*)$, with $H_0=w_*\bm1-M_w^{-1}M_0$ and $H_j=-M_w^{-1}M_j$ for $j=x,y,X$.  Writing $H=h_{\rm av}\bm1+\bm d\cdot\bm\sigma$, where $h_{\rm av}$ is the scalar band average and $\bm\sigma$ is the vector of Pauli matrices, the coefficients at the kinetic defect are
\begin{equation}
\bm d_0=\begin{pmatrix}
-2.10\!\times10^{-11}-3.85\!\times10^{-12}\ii\\
-2.13\!\times10^{-12}+1.07\!\times10^{-11}\ii\\
-1.387\!\times10^{-5}-7.440\!\times10^{-4}\ii
\end{pmatrix},
\end{equation}
and their Jacobian, with columns ordered as $(u_x,u_y,X-X_*)$, is
\begin{equation}
 \mathsf J_d=\begin{pmatrix}
-0.3501+0.0036\ii&0.0027-0.1786\ii&\sim10^{-12}\\
-0.0027+0.1786\ii&-0.3501+0.0036\ii&\sim10^{-11}\\
0&\sim10^{-11}&0.1520-0.0021\ii
\end{pmatrix}.
\label{S:Jd}
\end{equation}
For a $2\times2$ matrix the discriminant is $4\bm d\cdot\bm d$.  A Hermitian degeneracy requires three real Pauli components to vanish and is isolated in three real parameters.  An exceptional degeneracy imposes the two real conditions contained in one complex discriminant, producing a curve in $(u_x,u_y,X)$.  Rotational symmetry about $\bm B_0$ turns this curve into a circle.  Along $u_y=0$, $X=X_*$, the local equation $\bm d\cdot\bm d=0$ gives
$u_x\simeq\pm(2.470\times10^{-3}+1.3\times10^{-6}\ii)$.  The small imaginary part measures the error of the local frequency linearization.  Solving the full nonlinear conditions $\det\mathsf M^+=\partial_w\det\mathsf M^+=0$ with real $(u_\perp,X)$ gives the values quoted in the Letter; the determinant residual is of order $10^{-12}$.  The subscript ${\rm EP}$ below labels a point on this exceptional locus.

The defect is exceptional rather than a semisimple crossing.  The full $3\times3$ singular values are approximately $(1.98,1.05\times10^{-2},2.5\times10^{-11})$, so the geometric nullity is one at a double characteristic value.  Tracking both full nonlinear roots around
\begin{equation}
 u_x=u_{\perp,{\rm EP}}+5.0\times10^{-4}\cos\theta,\qquad
 X=X_{\rm EP}+1.3\times10^{-3}\sin\theta
\end{equation}
gives a direct root separation of about $2.0\times10^{-3}$ after one turn, while the crossed pairing agrees to $3\times10^{-14}$.  After two turns each root returns to itself with the same accuracy.  This is the square-root monodromy expected for an exceptional ring.

\section{Chern charge and causal reconstruction}

The enclosing ellipsoid has radii $u_r=1.8u_{\perp,{\rm EP}}\simeq0.00445$ and $X_r=0.055$.  At every vertex we solve the full nonlinear Schur determinant for both roots and verify them with the smallest singular value of the original $3\times3$ matrix.  The vertical reference line $\operatorname{Re}w=0.61779$ separates them everywhere.  The lower-real-part electric root therefore defines $\mathcal E_p=\ker\Dret[w(p),p]\subset\mathbb C^3$, even though the two roots exchange on a loop linking the exceptional ring inside the surface.

For an oriented triangular face $(ijk)$, let $|E_i^R\rangle$ and $\langle E_i^L|$ denote normalized right and left electric null vectors of the selected nonlinear root.  We use a gauge-invariant triangular link construction~\cite{Fukui2005}; related simplicial characteristic-class methods for phase-space polarization bundles are developed in Ref.~\cite{Bohlsen2026}.  The right-electric-root face flux is
\begin{equation}
 \Phi_{ijk}=\arg[\langle E_i^R|E_j^R\rangle\langle E_j^R|E_k^R\rangle\langle E_k^R|E_i^R\rangle],
 \qquad C^-_{{\rm F},R}=-\frac{1}{2\pi}\sum_{(ijk)}\Phi_{ijk}.
 \label{S:faure-discrete}
\end{equation}
The explicit minus sign converts the forward-link phase into Faure's convention $C_{\rm F}=(\ii/2\pi)\int d\langle E^R,dE^R\rangle$.  We repeat the unitary construction independently for the left electric root line bundle.  A one-sided biorthogonal link $\langle E_i^L|E_j^R\rangle/|\langle E_i^L|E_j^R\rangle|$ is unsuitable here because its value on a reversed edge is not generally the inverse.

\begin{table}[H]
\caption{Nonlinear root-bundle convergence.}
\begin{ruledtabular}
\begin{tabular}{rrrrr}
vertices&faces&$C^-_{{\rm F},R}$&$C^-_{{\rm F},L}$&min line gap\\\hline
12&20&$+1$&$+1$&$1.11\times10^{-3}$\\
42&80&$+1$&$+1$&$1.11\times10^{-3}$\\
162&320&$+1$&$+1$&$1.11\times10^{-3}$
\end{tabular}
\end{ruledtabular}
\end{table}
The nonlinear-root residual is below $10^{-13}$ on all three meshes.  To test inheritance from the cold Weyl point, we repeat the right- and left-root calculations on one fixed ellipsoid, with $(u_r,X_r)=(0.0048,0.055)$, throughout the temperature continuation.  This surface encloses the cold semisimple point and the exceptional ring at every later node.

\begin{table}[H]
\caption{Cold--hot bulk continuation.  Both electron temperatures are scaled at fixed $T_{e2}/T_{e1}=10$; the table lists $T_{e2}$.  The quantity $g$ is the minimum distance in $\operatorname{Re}w$ from either selected root to the moving vertical reference line.}
\begin{ruledtabular}
\begin{tabular}{rrrrr}
$T_{e2}$ (keV)&$u_{\perp,{\rm EP}}$&$C^-_{{\rm F},R}$&$C^-_{{\rm F},L}$&$g$\\\hline
0.01&0&$+1$&$+1$&$1.55\times10^{-3}$\\
1.0&0&$+1$&$+1$&$1.54\times10^{-3}$\\
3.0&$1.8\times10^{-6}$&$+1$&$+1$&$1.52\times10^{-3}$\\
5.0&$1.96\times10^{-4}$&$+1$&$+1$&$1.49\times10^{-3}$\\
8.0&$2.47\times10^{-3}$&$+1$&$+1$&$1.24\times10^{-3}$
\end{tabular}
\end{ruledtabular}
\end{table}

Across 11 temperature nodes from $T_{e2}=0.01$ to $8\,$keV, both electric-root bundles retain Chern number $+1$ and the enclosing surface remains line gapped.  The tracked defect stays within $51.5\%$ of the ellipsoid radius, the two-root separation exceeds $2.48\times10^{-3}$, and the largest root residual is $2.7\times10^{-12}$.  Representative mesh refinements reproduce the same integer, while the first retarded tensor differs from the exact cold tensor by $1.8\times10^{-6}$ in norm.  Thus the point-to-ring transition moves the defect without expelling charge through the enclosing surface on the sampled temperature path.  The stored data retain both $C^-_{\rm F}=+1$ and the raw forward-link sum $C_{\rm link}=-1$.

\subsection{Orientation and density-gradient sign}

We orient the phase-space surface by the outward orientation inherited from
\begin{equation}
 du_x\wedge du_y\wedge dX
 =du_y\wedge dX\wedge du_x,
 \label{S:orientation}
\end{equation}
use Faure's lower-line convention $C^-_{\rm F}=(\ii/2\pi)\int\mathcal F_R$, and quantize $u_x\mapsto-\ii\epsilon\partial_X$.  Here $\mathcal F_R=d\langle E^R,dE^R\rangle$ is the curvature two-form of the right electric root line bundle.  The forward-link formula approximates $-C^-_{\rm F}$, which explains the minus sign in Eq.~(\ref{S:faure-discrete}).  The Hermitian Dirac calibration
\begin{equation}
 h_{\rm D}=c_xu_x\sigma_x+c_yu_y\sigma_y+m'(X-X_c)\sigma_z
 \label{S:dirac-sign}
\end{equation}
uses real slopes $c_x$, $c_y$, and $m'=\partial_Xm|_{X_c}$.  It gives the phase-space chirality $\chi_X$ and the upward spectral flow $\operatorname{Sf}_{\uparrow}$ as
\begin{equation}
 \chi_X=\operatorname{sgn}(c_xc_ym'),\qquad
 C^-_{{\rm F},X}=\chi_X,\qquad
 \operatorname{Sf}_{\uparrow}=\chi_X=C^-_{{\rm F},X}.
 \label{S:index-sign}
\end{equation}
Equation~(\ref{S:index-sign}) fixes the orientation at the Hermitian endpoint.  The bulk cold--hot homotopy transports the charged root bundle, while a separate continuation at $u_y=0$ identifies the ancestry of the response pole.  The finite-window crossing is then evaluated directly at the hot endpoint.

The sign of the density gradient enters separately through the coordinate pullback.  If the homogeneous bulk charge is first defined with density itself as the control coordinate, with orientation $du_y\wedge dn\wedge du_x$, pullback by $n=n(X)$ gives
\begin{equation}
 C^-_{{\rm F},X}=\operatorname{sgn}\!\left(\frac{dn}{dX}\right)C^-_{{\rm F},n}.
 \label{S:density-pullback}
\end{equation}
For the equilibrium used here,
\begin{equation}
 \frac{dn}{dX}=\frac{n_+-n_-}{\sqrt\pi}e^{-X^2}<0,
\end{equation}
and hence $C^-_{{\rm F},X}=-C^-_{{\rm F},n}$.  Our numerical value is already the phase-space charge $C^-_{{\rm F},X}$, computed directly on the actual $(u_x,u_y,X)$ surface.  Denote the upward-minus-downward count in the tested window $\mathcal W$ by $\nu_{\mathcal W}$, as defined explicitly in Eq.~(\ref{S:finite-window-count}).  Consequently
\begin{equation}
 C^-_{{\rm F},X}=+1,\qquad C^-_{{\rm F},n}=-1,\qquad
 \nu_{\mathcal W}=+1\ \text{in the tested line-gap window}.
 \label{S:reported-signs}
\end{equation}
Reversing the density profile reverses the pullback from a density-coordinate bulk charge to the phase-space charge.  Because the reported $C^-_{{\rm F},X}$ is computed directly in $(u_x,u_y,X)$, no additional gradient sign is applied.  Faure's theorem fixes the cold Hermitian convention.  Equation~(\ref{S:reported-signs}) is a finite-window equality at the hot endpoint; the calculation does not establish index invariance over the full temperature--$u_y$ plane.

\subsection{Electric/kinetic-response bundle isomorphism}

Auxiliary-state descriptions restore a frequency-linear enlarged problem in dispersive electromagnetism~\cite{RamanFan2010,Silveirinha2015}.  In kinetic theory, reduction to electromagnetic variables also encounters the homogeneous free-streaming sector.  The following proposition states the conditions under which the electric reduction preserves the selected kinetic topology.

At fixed $X$, let
\begin{equation}
 f_{0s}^{(0)}(X,\bm v)=\sum_{a=1}^{2}n_{sa}(X)
 (\pi v_{tsa}^2)^{-3/2}e^{-v^2/v_{tsa}^2}
 \label{S:frozen-equilibrium}
\end{equation}
be the frozen isotropic equilibrium that enters the principal symbol.  The Fourier-transformed linearized Vlasov equation can then be written
\begin{equation}
 \mathcal L_s(z,\bm k;X)\delta f_s
 =\mathcal S_s(z,\bm k;X)\bm E,
 \label{S:vlasov-operator}
\end{equation}
where
\begin{align}
 \mathcal L_s
 &=-\ii z+\ii\bm k\cdot\bm v
 +\frac{q_s}{m_s}(\bm v\times\bm B_0)\cdot\nabla_{\bm v},\\
 \mathcal S_s\bm E
 &=-\frac{q_s}{m_s}
 \left[\bm E+\bm v\times\frac{\bm k\times\bm E}{z}\right]
 \cdot\nabla_{\bm v}f_{0s}^{(0)}(X,\bm v).
\end{align}

We now specify the velocity-space meaning of the continued response.  For
$a>0$, let $\mathcal X_s^{(a)}$ be the space of Schwartz functions on
$\mathbb R^3_{\bm v}$ that extend holomorphically in $v_z$ to
$|\operatorname{Im}v_z|<a$.  Its Fr\'echet topology is generated by
\begin{equation}
 q_{N,\alpha,a'}(g)=
 \sup_{|\eta|\le a'}\sup_{\bm v\in\mathbb R^3}
 (1+|\bm v|)^N
 \left|\partial_{\bm v}^{\alpha}
 g(\bm v+\ii\eta\hat{\bm z})\right|,
 \qquad 0<a'<a,
 \label{S:analytic-test-seminorm}
\end{equation}
for every $N$ and multi-index $\alpha$.  We use the rigged space
\begin{equation}
 \mathcal X_s^{(a)}\subset
 \mathcal H_s=L^2(\mathbb R^3,d^3v)\subset
 \mathcal X_s^{(a)\prime},
 \label{S:rigged-velocity-space}
\end{equation}
where $\mathcal X_s^{(a)\prime}$ is the strong dual.  Its seminorms are
$Q_B(F)=\sup_{g\in B}|\langle F,g\rangle|$, with $B$ any bounded subset of
$\mathcal X_s^{(a)}$.  This is the standard rigged-space interpretation of
an analytically continued resonance~\cite{Chiba2015,PrigoginePetrosky1998}.  The Maxwellian source
$\mathcal S_s\bm E$ belongs to $\mathcal X_s^{(a)}$ for every finite $a$.

We define the retarded response operator by
\begin{equation}
 \mathcal R_s^+(z,\bm k;X)=\bigl[\mathcal L_s(z,\bm k;X)\bigr]_{\rm ret}^{-1},
 \label{S:retarded-resolvent}
\end{equation}
where the subscript denotes continuation from $\operatorname{Im}z>0$ as a
map $\mathcal X_s^{(a)}\to\mathcal X_s^{(a)\prime}$.  Thus
$\mathcal A_s^+=\mathcal R_s^+\mathcal S_s$ belongs to
$\mathcal L(\mathbb C^3,\mathcal X_s^{(a)\prime})$.  After continuation,
$\delta f_s=\mathcal A_s^+\bm E$ is an analytic functional, or resonant
state, rather than an $L^2$ eigenvector of the conservative Vlasov generator.

The retarded sheet can be chosen uniformly on the enclosing surface used in
the Letter.  Here $k_z>0$ is fixed, $z(p)$ is continuous, and $\Sigma$ is
compact.  Define
\begin{equation}
 M_\Sigma=\max\!\left\{0,
 \sup_{p\in\Sigma}\left[-\frac{\operatorname{Im}z(p)}{k_z}\right]
 \right\}<\infty.
 \label{S:contour-depth-bound}
\end{equation}
The Maxwellian source is entire in $v_z$, so the strip width may be chosen
with $a>M_\Sigma$.  Select
\begin{equation}
 M_\Sigma<\eta<a,
 \qquad \mathcal C_\eta=\mathbb R-\ii\eta.
 \label{S:uniform-landau-contour}
\end{equation}
This contour lies below every resonance pole
\begin{equation}
 v_{s\ell}(p)=\frac{z(p)-\ell\Omega_s}{k_z},
 \qquad p\in\Sigma,\quad \ell\in\mathbb Z.
 \label{S:resonance-poles}
\end{equation}
Indeed, $\operatorname{Im}v_{s\ell}=\operatorname{Im}z/k_z$ is independent
of $s$ and $\ell$.  For $v_z=t-\ii\eta$ on $\mathcal C_\eta$,
\begin{equation}
 \left|z(p)-k_zv_z-\ell\Omega_s\right|
 \ge k_z(\eta-M_\Sigma)>0,
 \label{S:uniform-resonance-separation}
\end{equation}
so every resonant denominator remains a uniform positive distance from zero.
Write $\delta_\Sigma=k_z(\eta-M_\Sigma)>0$.  For real $\lambda$, the standard
Bessel identities
\[
 \sum_{\ell\in\mathbb Z}J_\ell^2(\lambda)=1,
 \qquad
 \sum_{\ell\in\mathbb Z}[J_\ell'(\lambda)]^2=\frac12,
 \qquad
 \sum_{\ell\in\mathbb Z}
 \left[\frac{\ell J_\ell(\lambda)}{\lambda}\right]^2=\frac12
\]
hold, with the zero-argument ratios understood by their analytic limits.
Cauchy--Schwarz therefore bounds every absolute Bessel product in the response
kernel uniformly on $\Sigma$.  After pairing with a bounded set
$B\subset\mathcal X_s^{(a)}$, the remaining polynomial velocity factors are
bounded by an analytic-Schwartz seminorm, while the shifted Maxwellian supplies
the integrable majorant
$C_{B,\Sigma}\delta_\Sigma^{-1}(1+|t|+v_\perp)^m
\exp[-(t^2+v_\perp^2)/v_{tsa}^2]$ on $v_z=t-\ii\eta$, where $m$ is finite
and $C_{B,\Sigma}$ is independent of $p\in\Sigma$.
Thus the Bessel-weighted harmonic series and Gaussian velocity integrals
converge locally uniformly in $p$.  Dominated
convergence makes $p\mapsto\mathcal A_s^+(p)$ continuous in the induced
strong-dual operator topology.  This means uniform convergence on bounded
electric-field sets, measured by the strong-dual seminorms:
\begin{equation}
 \sup_{\|\bm E\|\le1}\sup_{g\in B}
 \left|
 \left\langle
 [\mathcal A_s^+(p)-\mathcal A_s^+(p_0)]\bm E,g
 \right\rangle\right|\longrightarrow0
 \quad (p\to p_0)
 \label{S:strong-operator-continuity}
\end{equation}
for every bounded $B\subset\mathcal X_s^{(a)}$.  The fixed contour is in the
retarded Landau homology class obtained from $\operatorname{Im}z>0$ and fixes
one causal continuation over all of $\Sigma$~\cite{FriedConte1961}.  This is
the precise continuity used below.  The range $\mathcal A_s^+\bm E$ retains
Gaussian decay along $\mathcal C_\eta$, so its charge and current moments are
absolutely convergent and continuous in $p$.

\emph{Proposition.---}
Let $p=(u_x,u_y,X)\in\Sigma$, with $u_z$ fixed, and let $w(p)$ be a selected simple, line-gap-isolated zero of $\Dret$ on the closed surface $\Sigma$.  Set $z(p)=\omega_{ce}w(p)\neq0$ and use the continuous retarded maps $\mathcal A_s^+(p)$ defined above.  Let $\mathcal E\to\Sigma$ be the electric root line bundle.  In the fixed ambient space
\begin{equation}
 \mathcal Y=\mathbb C^3_{\bm E}\oplus\mathbb C^3_{\bm B}
 \oplus\bigoplus_s\mathcal X_s^{(a)\prime},
 \label{S:response-ambient-space}
\end{equation}
with its product topology, let $\mathcal K^{\rm ret}\to\Sigma$ be the selected retarded kinetic-response line bundle whose fiber consists of $(\bm E,\bm B,\{\delta f_s\}_s)$ after imposing the retarded condition.  Equivalently, $\mathcal K^{\rm ret}$ may be defined after quotienting the free-streaming homogeneous kernel.  Then electric projection is a complex line-bundle isomorphism
\begin{equation}
 \pi_E:\mathcal K^{\rm ret}\xrightarrow{\ \simeq\ }\mathcal E,
 \qquad
 c_1(\mathcal K^{\rm ret})=c_1(\mathcal E).
 \label{S:electric-full-isomorphism}
\end{equation}

To prove the statement, Eq.~(\ref{S:vlasov-operator}) gives the general Vlasov solution at $p\in\Sigma$ as
\begin{equation}
 \delta f_s=\mathcal A_s^+(p)\bm E+h_s,
 \qquad h_s\in\ker\mathcal L_s(p).
 \label{S:vlasov-general-solution}
\end{equation}
The retarded prescription fixes $h_s=0$; the same result follows in the quotient by the free-streaming kernel~\cite{VanKampen1955,Case1959,Ramos2019}.  Faraday's law gives $\bm B=\bm k(p)\times\bm E/z(p)$, so define
\begin{equation}
 G_p:\bm E\longmapsto
 \left(\bm E,\frac{\bm k(p)\times\bm E}{z(p)},
 \{\mathcal A_s^+(p)\bm E\}_s\right).
 \label{S:response-graph-map}
\end{equation}
Equations~(\ref{S:strong-operator-continuity}) and
(\ref{S:response-ambient-space}) make $G_p$ a continuous linear map into the
selected retarded response fiber.  Its first component gives
$\pi_EG_p=1$.  Conversely, every solution in $\mathcal K^{\rm ret}$ satisfies
Eq.~(\ref{S:vlasov-general-solution}) with $h_s=0$ and is therefore generated
by its electric component.  Hence $G_p\pi_E=1$ on the selected response
fiber, proving Eq.~(\ref{S:electric-full-isomorphism}).

The parameter-dependent map $G_p$ need not be unitary, so the electric and reconstructed-response Berry curvatures can differ pointwise.  Their first Chern classes nevertheless agree.  The calculation verifies simple line-gap isolation and continuous tracking of the selected electric root on $\Sigma$.  Equations~(\ref{S:analytic-test-seminorm})--(\ref{S:strong-operator-continuity}) specify the topology and the global retarded sheet for the Maxwellian principal symbol.  Quotienting removes arbitrary $h_s$, and therefore the homogeneous $\bm E=0$ sector, by definition; it does not assert that the unquotiented free-streaming continuum is topologically trivial.

\section{Interface resonance pole and finite-window oriented crossing count}

The local calculation predicts the branch followed by the Weyl-quantized principal-symbol response below.  A complex-frequency zero of the retarded boundary operator is a pole of its Green function.  The kinetic continuation of the cold edge eigenmode is therefore sought as a localized pole with directed real dispersion and resonant decay.  The frequency-linearized operator is
\begin{equation}
 \widehat H=H_0+H_yu_y+H_X\xi-\ii\epsilon H_x\partial_\xi,
 \qquad \xi=X-X_{\rm EP},\quad \epsilon=\frac{c}{\omega_{ce}L}.
\end{equation}
Substitution of $\psi=e^{-\alpha\xi^2/2+\beta\xi}\chi$, where $\alpha,\beta\in\mathbb C$ are Gaussian localization parameters and $\chi$ is a constant two-component spinor, separates powers of $\xi$:
\begin{align}
 (H_X+\ii\epsilon\alpha H_x)\chi&=0,\label{S:alpha}\\
 (H_0+H_yu_y-\ii\epsilon\beta H_x)\chi&=w\chi.\label{S:beta}
\end{align}
Equation (\ref{S:alpha}) is a quadratic equation in $\alpha$ and has exactly one root with positive real part,
\begin{equation}
 \alpha=28.12-0.037\ii.
\end{equation}
At $u_y=0$, $\beta\simeq8.4\times10^{-4}+0.138\ii$, giving
$w\simeq0.61780-1.040\times10^{-3}\ii$.  The intensity has
\begin{equation}
 \langle X\rangle=X_{\rm EP}+\frac{\operatorname{Re}\beta}{\operatorname{Re}\alpha}\simeq0.2222,
 \qquad \sigma_X=(2\operatorname{Re}\alpha)^{-1/2}\simeq0.1333.
\end{equation}
Equations (\ref{S:alpha})--(\ref{S:beta}) give a branch exactly linear in $u_y$.

As an independent full-line calculation, we expand the mode in an orthonormal Hermite basis with oscillator length $(\operatorname{Re}\alpha)^{-1/2}$.  From 12 to 36 basis functions, the low-order state agrees with the analytic frequency to about $10^{-14}$ and with its width to better than $10^{-9}$.  A high-order cutoff state broadens under refinement and is identified as spectral pollution.

\subsection{Full-symbol Weyl quantization and Keldysh pole}

The local Dirac result is analytic, but for $L=0.1\,$m its momentum width is not confined to the strictly linear neighborhood of the Weyl exceptional ring (WER).  We therefore quantize the full nonlinear $3\times3$ retarded principal symbol.  On a uniform periodic grid $\{X_i\}_{i=0}^{N_X-1}\subset[X_{\min},X_{\max})=[-1.25,1.75)$, let $q_m$ be the discrete Fourier wave numbers and let $\mu,\nu\in\{1,2,3\}$ label electric-field components.  The Weyl-quantized matrix operator is
\begin{equation}
 [\Op\Dret(w)]_{i\mu,j\nu}=\frac1{N_X}\sum_m e^{\ii q_m(X_i-X_j)}
\bigl(\Dret\bigr)_{\mu\nu}\!\left(w,\epsilon q_m,u_y,\frac{X_i+X_j}{2}\right).
 \label{S:weyl}
\end{equation}
For a normalized right electric null vector with grid components $E^R_{i\mu}$, define its spatial intensity, centroid, and root-mean-square width by
\begin{equation}
 \mathcal I_i=\sum_{\mu=1}^{3}|E^R_{i\mu}|^2,
 \qquad \sum_i\mathcal I_i=1,
 \qquad \langle X\rangle=\sum_i\mathcal I_iX_i,
 \qquad \sigma_X^2=\sum_i\mathcal I_i(X_i-\langle X\rangle)^2.
 \label{S:localization-measures}
\end{equation}
We quantify localization at the physical interface by
\begin{equation}
 P_{\rm int}=\sum_{|X_i-X_{\rm EP}|<0.25}\mathcal I_i,
 \qquad
 P_{\rm seam}=\sum_{\substack{X_i<X_{\min}+0.15L_{\rm box}\\
                         \text{or }X_i>X_{\max}-0.15L_{\rm box}}}\mathcal I_i,
 \quad L_{\rm box}=X_{\max}-X_{\min},
 \label{S:interface-seam-weights}
\end{equation}
where $P_{\rm seam}$ measures the intensity in the outer $15\%$ of the interval on each side of the artificial periodic seam.
The susceptibility is affine in each Maxwellian component density and hence affine in the total density for Eq.~(\ref{S:split}); only two endpoint symbols are needed for every Fourier momentum.  The periodic seam represents a second, remote interface.  For the central mode its weight is below $5\times10^{-4}$ at $N_X=16$ and decreases under refinement.

At each trial $w$, we compute the matrix eigenvalue closest to zero and its left and right vectors.  A two-real-variable Newton iteration sets this complex eigenvalue to zero.  The procedure solves a nonlinear response-root problem in the sense of Keldysh~\cite{Beyn2012}; it does not diagonalize a frequency-frozen Hamiltonian or the conservative Vlasov generator.

The word ``pole'' can be made precise.  Put $\mathcal T(w)=\Op\Dret(w)$ and normalize the null vectors at $w_0$ by $\|R\|_2=1$ and $\langle L|R\rangle=1$.  If
\begin{equation}
 \kappa=\langle L|\mathcal T'(w_0)|R\rangle\ne0,
 \label{S:keldysh-denominator}
\end{equation}
then Keldysh's local expansion gives
\begin{equation}
 \mathcal T(w)^{-1}=\frac{|R\rangle\langle L|}{(w-w_0)\kappa}+O(1).
 \label{S:keldysh-residue}
\end{equation}
For the $N_X=24$ Weyl-ordered pole, centered differences with steps $2\times10^{-5}$, $10^{-5}$, and $5\times10^{-6}$ give
\begin{equation}
 \kappa=6.2186+0.1681\ii,\qquad
 \left\|\mathop{\rm Res}_{w_0}\mathcal T^{-1}\right\|_F=0.16087.
\end{equation}
The last two $\kappa$ values differ relatively by $1.0\times10^{-9}$.  The zero is therefore algebraically simple, with a step-converged rank-one Green-function residue.  This residue distinguishes the response pole from a small singular value sampled at one frequency.

\begin{table}[H]
\caption{Convergence of the Weyl-quantized full nonlinear principal-symbol response at $u_y=0$.}
\begin{ruledtabular}
\begin{tabular}{rrrr}
$N_X$&$\operatorname{Re}w$&$10^3\operatorname{Im}w$&$P_{\rm seam}$\\\hline
16&0.6184659&$-1.05608$&$4.7\times10^{-4}$\\
20&0.6184794&$-1.05807$&$3.2\times10^{-6}$\\
24&0.6184795&$-1.05808$&$7.9\times10^{-9}$\\
32&0.6184795&$-1.05808$&$4.2\times10^{-13}$
\end{tabular}
\end{ruledtabular}
\end{table}

For all entries, the modulus $|\tau_0|$ of the matrix eigenvalue followed to
zero is below $6\times10^{-14}$.  Replacing the midpoint in Eq.~(\ref{S:weyl}) by the left coordinate gives the Kohn--Nirenberg ordering.  At $N_X=32$ the two orderings differ in $w$ by about $10^{-9}$, well below the physical precision quoted in the Letter.

To separate Fourier refinement from finite-box error, we keep the grid spacing $\Delta X=0.125$ fixed and enlarge the interval while increasing $N_X$ proportionally:
\begin{table}[H]
\caption{Box-length convergence at $u_y=0$ (Weyl ordering).}
\begin{ruledtabular}
\begin{tabular}{rrrr}
$X_{\max}-X_{\min}$&$N_X$&$\operatorname{Re}w$&$10^3\operatorname{Im}w$\\\hline
3.0&24&0.6184795223&$-1.058080335$\\
4.5&36&0.6184795223&$-1.058080336$\\
6.0&48&0.6184795223&$-1.058080337$
\end{tabular}
\end{ruledtabular}
\end{table}
The change from the shortest to the longest box is $4\times10^{-11}$ in complex frequency, and the seam intensity remains below $8\times10^{-9}$.  Across the same boxes, $|\kappa|=6.2209$ and the residue norm is $0.16087$ at the displayed precision.

\subsection{Cold-to-hot pole continuation and finite-window count}

To identify the hot pole as the continuation of the cold interface mode, we follow the Weyl-quantized nonlinear operator at $u_y=0$ through 12 temperature nodes.  Both electron components are scaled together at fixed $T_{e2}/T_{e1}=10$.  The zero-temperature endpoint uses the exact cold Maxwell--plasma tensor, while every positive-temperature node uses the retarded Vlasov susceptibility.

At every node we solve the nonlinear operator equation, track the right null vector by maximum overlap, and recompute the Keldysh denominator.  Representative $N_X=24$ results are
\begin{table}[H]
\caption{Boundary continuation of the Weyl-quantized full nonlinear principal symbol at $u_y=0$.  The cold row is an eigenmode of the Hermitian boundary operator; the finite-temperature rows are roots of the retarded response.}
\begin{ruledtabular}
\begin{tabular}{rrrr}
$T_{e2}$ (keV)&$\operatorname{Re}w$&$10^3\operatorname{Im}w$&$\sigma_X$\\\hline
0&0.618777&0&0.134\\
1&0.618840&0&0.134\\
3&0.618906&$-0.001$&0.134\\
5&0.618784&$-0.083$&0.134\\
8&0.618480&$-1.058$&0.135
\end{tabular}
\end{ruledtabular}
\end{table}
All 12 roots have nonzero, step-converged Keldysh denominators.  The smallest adjacent right-vector overlap is $0.9981$ at $N_X=24$, while $P_{\rm int}>0.933$ and $P_{\rm seam}<8\times10^{-9}$.  Increasing to $N_X=32$ changes the complex frequency by less than $2\times10^{-10}$ at every node.  The hot pole is therefore the kinetic continuation of the cold edge eigenmode along the sampled $u_y=0$ path.

\begin{table}[H]
\caption{Chiral edge branch obtained from the full nonlinear principal symbol ($N_X=24$, Weyl ordering).}
\begin{ruledtabular}
\begin{tabular}{rrrrr}
$u_y$&$\operatorname{Re}w$&$10^3\operatorname{Im}w$&$\langle X\rangle$&$\sigma_X$\\\hline
$-0.006$&0.616768&$-1.016$&0.223&0.135\\
$0$&0.618480&$-1.058$&0.227&0.135\\
$0.006$&0.620196&$-1.102$&0.231&0.135
\end{tabular}
\end{ruledtabular}
\end{table}

Extending the full nonlinear principal-symbol continuation to 25 equally spaced points in the finite line-gap window $\mathcal W=[-0.009,0.009]$ gives
\begin{equation}
 \frac{dw_{\rm edge}}{du_y}=0.2857-0.00713\ii
\end{equation}
from a linear fit.  On $\mathcal W$, every stored continuation root converges, $P_{\rm int}>0.929$, and $P_{\rm seam}<1.5\times10^{-8}$.  Pointwise Keldysh denominators were not recomputed in the $N_X=24,32$ scan files; simplicity across sampled momenta is instead checked in the sampled contour pole inventory below.  We define only the finite-window oriented count
\begin{equation}
 \nu_{\mathcal W}=\sum_{\substack{u_y^*\in\mathcal W:\\
 \operatorname{Re}w_{\rm edge}(u_y^*)=\operatorname{Re}w_{\rm EP}\\
 \partial_{u_y}\operatorname{Re}w_{\rm edge}(u_y^*)\ne0}}
 \operatorname{sgn}\!\left[
 \partial_{u_y}\operatorname{Re}w_{\rm edge}(u_y^*)\right].
 \label{S:finite-window-count}
\end{equation}
There is one upward crossing at $u_y=-2.420\times10^{-3}$ and no downward crossing.  At the hot endpoint, the directly computed values match:
\begin{equation}
 \boxed{\nu_{\mathcal W}=C^-_{\rm F}=+1}
 \label{S:kinetic-index-equality}
\end{equation}
on the stated line-gap window.  The largest zero-eigenvalue residual is $1.4\times10^{-13}$.  Faure proves the equality at the Hermitian phase-space endpoint, and Qin and Fu apply it to the cold-plasma topological Langmuir--cyclotron wave (TLCW)~\cite{Faure2023,QinFu2023}.  Equation~(\ref{S:kinetic-index-equality}) is the corresponding numerical equality at the hot endpoint.  It is local to the declared phase-space and spectral window.

The crossing itself converges with Fourier order:
\begin{table}[H]
\caption{Finite-window crossing convergence.  ``validated'' counts continuation points that pass the localization and seam tests.}
\begin{ruledtabular}
\begin{tabular}{rrrr}
$N_X$&validated&$u_y^*$&$\nu_{\mathcal W}$\\\hline
16&20/25&$-2.377\times10^{-3}$&$+1$\\
24&25/25&$-2.420056\times10^{-3}$&$+1$\\
32&25/25&$-2.420057\times10^{-3}$&$+1$
\end{tabular}
\end{ruledtabular}
\end{table}
At $N_X=16$ the coarse grid reaches the seam threshold on the far positive side after the crossing.  We retain it as a low-resolution diagnostic.  The $N_X=24$ and $N_X=32$ scans validate every point and agree in $u_y^*$ to $5\times10^{-10}$.

To check that continuation did not miss competing poles, we also applied the Beyn contour-moment method to the full $N_X=24$ nonlinear operator at $u_y=0$.  On the ellipse centered at $0.618-1.5\times10^{-3}\ii$ with radii $(0.042,0.003)$, the zeroth contour moment has numerical rank four and all four candidates refine to operator zeros:
\begin{table}[H]
\caption{Broad-window nonlinear spectrum at $u_y=0$.}
\begin{ruledtabular}
\begin{tabular}{rrlrr}
$\operatorname{Re}w$&$10^3\operatorname{Im}w$&localization&$\sigma_X$&$P_{\rm seam}$\\\hline
0.582317&$-0.448$&central bound pole&0.186&$2.3\times10^{-8}$\\
0.618480&$-1.058$&charged pole&0.135&$7.9\times10^{-9}$\\
0.636011&$-1.633$&periodic-seam pole&1.137&0.647\\
0.654401&$-2.625$&central bound pole&0.198&$7.7\times10^{-8}$
\end{tabular}
\end{ruledtabular}
\end{table}
The contour stays nonsingular, with minimum singular value $6.6\times10^{-3}$ on its nodes.  The neighboring central poles are converged bound states.  The broad pole at $0.636011-1.633\times10^{-3}\ii$ moves with Fourier size and accumulates at the remote periodic seam.  Localization alone therefore does not identify the topological branch.

We search the full ellipse at five momenta $u_y=-0.009,-0.0045,0,0.0045,0.009$ using the lower-cost $N_X=16$ discretization.  At every momentum the Beyn moment has rank four, four candidates refine to zeros, the contour minimum singular value exceeds $6.5\times10^{-3}$, and every zero has a nonzero step-converged Keldysh denominator.  Continuous root matching gives
\begin{table}[H]
\caption{Sampled contour pole inventory across the finite line-gap window.  The ranges are over the five sampled $u_y$ values.}
\begin{ruledtabular}
\begin{tabular}{lllcc}
branch&$\operatorname{Re}w$ range&classification&crosses line&simple\\\hline
lower&0.58217--0.58230&central bound&no&yes\\
charged&0.61590--0.62103&central interface&upward once&yes\\
finite-box&0.62589--0.63094&remote/seam related&no&yes\\
upper&0.65443--0.65457&central bound&no&yes
\end{tabular}
\end{ruledtabular}
\end{table}
The five-momentum contour inventory checks for competing crossings inside the declared frequency ellipse.  The $N_X\in\{24,32\}$ continuation instead establishes convergence and localization of the charged pole.  Only the branch continued from the charged defect crosses the reference line in $\mathcal W$.

\section{Numerical robustness beyond the baseline model}

Gauss--Laguerre orders 12--32 all give
$w_{\rm EP}\simeq0.61779-8.07\times10^{-4}\ii$,
$u_{\perp,\rm EP}\simeq2.470\times10^{-3}$, and $X_{\rm EP}\simeq0.22215$.  The spread in $u_{\perp,\rm EP}$ is $3\times10^{-11}$ and the other coordinates vary still less.

The main calculation neglects ion susceptibility because $z\simeq0.618\omega_{ce}\gg\Omega_i$.  Restoring both proton Maxwellians gives $w_{\rm EP}\simeq0.61787-8.08\times10^{-4}\ii$, $u_{\perp,\rm EP}\simeq0.002476$, and $X_{\rm EP}\simeq0.2225$.  The root line bundles still have $C^-_{{\rm F},R}=C^-_{{\rm F},L}=+1$ with a line-gap margin of $1.1\times10^{-3}$, and the edge pole moves only to $w\simeq0.61856-1.06\times10^{-3}\ii$.

Keeping the bulk densities fixed while changing the erf width to $0.8L$ and $1.2L$ gives edge poles $0.61865-1.063\times10^{-3}\ii$ and $0.61837-1.055\times10^{-3}\ii$, with widths $0.122L$ and $0.148L$, respectively.  The pole therefore shifts smoothly but remains localized.

\subsection{Relativistic principal-symbol response}

The hotter electron component has $\Theta_{e2}=T_{e2}/m_ec^2=1.57\times10^{-2}$.  This is small for bulk thermodynamic moments, although resonant denominators can amplify relativistic corrections.  We therefore repeat the bulk and boundary calculations with a self-consistent relativistic equilibrium and susceptibility.

In kinetic momentum $\bm p$, the relativistic Vlasov equation is
\begin{equation}
 \partial_t f_s+\bm v_s\cdot\nabla_{\bm x}f_s
 +q_s(\bm E+\bm v_s\times\bm B)\cdot\nabla_{\bm p}f_s=0,
 \quad
 \bm v_s=\frac{\bm p}{\gamma_sm_s},\quad
 \gamma_s=\sqrt{1+\frac{p^2}{m_s^2c^2}}.
 \label{S:rel-vlasov}
\end{equation}
For uniform $\bm B_0=B_0\hat{\bm z}$, the invariant replacing
$x+v_y/\Omega_s$ is
\begin{equation}
 \mathcal X_s^{\rm rel}=x+\frac{p_y}{q_sB_0}
 =x+\frac{\gamma_sv_y}{\Omega_{0s}},
 \qquad \Omega_{0s}=\frac{q_sB_0}{m_s}.
 \label{S:rel-invariant}
\end{equation}
Together with the conserved energy, this gives the exact stationary
equilibrium
\begin{align}
 f_{0s}^{\rm rel}(x,\bm p)
 &=\sum_aN_{sa}(\mathcal X_s^{\rm rel})F_{J,sa}(\bm p),\label{S:rel-equilibrium}\\
 F_{J,sa}(\bm p)
 &=\frac{\exp(-\gamma_s/\Theta_{sa})}
 {4\pi m_s^3c^3\Theta_{sa}K_2(1/\Theta_{sa})},
 \qquad \Theta_{sa}=\frac{T_{sa}}{m_sc^2},
\end{align}
where $K_2$ is a modified Bessel function and $F_{J,sa}$ is normalized in
momentum space~\cite{Juttner1911}.  Because
$\partial_{p_y}F_{J,sa}=-(v_y/T_{sa})F_{J,sa}$, integration by parts gives
\begin{equation}
 j_{y,sa}=q_s\int v_y f_{0sa}^{\rm rel}\,d^3p
 =\frac{T_{sa}}{B_0}\frac{dn_{sa}}{dx}.
 \label{S:rel-current}
\end{equation}
The same constant-pressure split in Eq.~(\ref{S:split}) therefore cancels the
total equilibrium current, and equal electron and ion densities cancel the
charge.  Equations (\ref{S:rel-equilibrium}) and (\ref{S:rel-current}) thus
define an exact Maxwell-self-consistent relativistic equilibrium with uniform
$\bm B_0$ and $\bm E_0=0$.

The physical component density is now the orbit convolution
\begin{equation}
 n_{sa}(x)=\int N_{sa}\!\left(x+\frac{p_y}{q_sB_0}\right)
 F_{J,sa}(\bm p)\,d^3p,
 \label{S:rel-convolution}
\end{equation}
whose kernel is not Gaussian.  We invert Eq.~(\ref{S:rel-convolution}) in
Fourier space without spectral filtering.  For the stated density and
temperature profiles, both recovered guiding-center densities remain
positive.  The rms orbit displacement of the $8\,$keV electron component is
$2.17\times10^{-3}L$, close to but distinct from its nonrelativistic rms
orbit-width estimate.

For the frozen principal symbol, we specialize the relativistic gyrotropic
Vlasov susceptibility to the isotropic Maxwell--J\"uttner distribution using
the momentum-space representation of Refs.~\cite{Verscharen2018,YoonDavidson1990}.
The resonance denominator is
\begin{equation}
 z-k_\parallel v_\parallel-\ell\Omega_{0s}/\gamma_s,
 \qquad \ell\in\mathbb Z,
 \label{S:rel-resonance}
\end{equation}
so particles of different momentum sample different gyrofrequencies.  For damped roots, the parallel-momentum contour passes below the continued Landau and cyclotron poles while avoiding the branch points of $\gamma_s$.  Changing among admissible retarded contours leaves the roots invariant within quadrature error.
The relativistic tensor approaches the cold tensor with relative error
$3.4\times10^{-7}$ in the cold-limit test.  Radial-by-parallel quadratures from
$20\times32$ to $40\times72$, together with several contour depths, give the
same exceptional root to the digits reported below.

At $u_z=1$, the relativistic nonlinear double-root conditions give
\begin{equation}
 w_{\rm EP}^{\rm rel}=0.611932-2.188\times10^{-3}\ii,\qquad
 u_{\perp,{\rm EP}}^{\rm rel}=7.255\times10^{-3},\qquad
 X_{\rm EP}^{\rm rel}=0.22114.
 \label{S:rel-ep}
\end{equation}
The algebraically double root has a one-dimensional null space.  A loop
linking the ring exchanges the two roots after one circuit and returns each
root after two; the largest root residual along the tracked loop is
$6.9\times10^{-16}$.  On an enclosing ellipsoid with 162 vertices and 320
faces, the right and left nonlinear root bundles both give
$C^-_{{\rm F},R}=C^-_{{\rm F},L}=+1$, and the minimum line-gap margin is
$3.1\times10^{-3}$.

To test whether the integer is inherited from the nonrelativistic response,
we use the diagnostic homotopy
\begin{equation}
 \mathsf D_\tau^+=(1-\tau)\mathsf D_{\rm nr}^+
 +\tau\mathsf D_{\rm rel}^+,
 \qquad 0\leq\tau\leq1.
 \label{S:rel-homotopy}
\end{equation}
At nine sampled values of $\tau$, the same fixed 42-vertex ellipsoid encloses the defect, remains line gapped, and carries $C^-_{\rm F}=+1$.  The smallest margin is $3.6\times10^{-3}$.  This is a mesh-resolved numerical homotopy rather than an interval-arithmetic proof between nodes.

Weyl quantization of the relativistic nonlinear principal symbol gives the
simple Keldysh pole
\begin{equation}
 w_{\rm edge}^{\rm rel}\simeq0.6126641-2.9826\times10^{-3}\ii,
 \qquad \sigma_X=0.1362.
 \label{S:rel-edge}
\end{equation}
The $N_X=24$ and $N_X=32$ Weyl-ordered results agree at the displayed
accuracy, and Kohn--Nirenberg ordering gives the same pole.  A 25-point scan
over $-0.009\leq u_y\leq0.009$ contains one upward crossing and no downward
crossing, with
$\partial w/\partial u_y=0.28297-0.01570\ii$.  Throughout the scan the
central-interface weight exceeds $0.927$, whereas the artificial-seam weight
is below $1.7\times10^{-8}$.  Relative to the nonrelativistic calculation, the
real EP frequency and the real group velocity change by about $1\%$, the ring
radius increases by a factor of $2.9$, and the edge-pole damping increases by
a factor of $2.8$.

The graph-map argument in Eqs.~(\ref{S:electric-full-isomorphism})--(\ref{S:response-graph-map}) extends after replacing $\mathcal R_s^+$ and $\mathcal S_s$ by their relativistic counterparts.  Under the same spectral assumptions, electric projection remains the inverse graph map and the two bundles have the same first Chern class.  The computed enclosing surface stays line gapped; the homogeneous $\bm E=0$ sector is excluded by the same retarded quotient construction.

Relativistic kinematics preserves the charged continuation and crossing direction for the tested family, while substantially changing the lifetime.  The next calculation tests the leading finite-orbit correction to this retarded response.

\subsection{Orbit-resolved nonuniform retarded-response benchmark}

We construct the linear retarded response of Eq.~(\ref{S:rel-equilibrium}) by integrating the perturbed distribution along exact unperturbed relativistic gyro-orbits.  The kernel includes the electromagnetic Lorentz force, Maxwell--J\"uttner momentum integration, the guiding-center gradient source, and finite-orbit sampling.  It is a nonlocal linear-response calculation rather than a local dielectric approximation.

Expand the electric field in Fourier modes with
$\bm k_n=(k_{x,n},k_y,k_z)$.  Faraday's law gives
$\bm B_n=\bm k_n\times\bm E_n/z$.  For each input mode, exact propagation
along the orbit and retarded elimination of momentum space produce a generally
non-diagonal conductivity block $\bm\sigma^+_{mn}(z)$, defined by
$\bm J_m=\sum_n\bm\sigma^+_{mn}(z)\bm E_n$.  The electric-field
operator is
\begin{equation}
 \mathsf T^+_{{\rm orbit},mn}(z)
 =\delta_{mn}\!\left[\bm1-n_n^2\bm1+\bm n_n\bm n_n^{\mathsf T}\right]
 +\frac{\ii\bm\sigma^+_{mn}(z)}{\epsilon_0z},
 \qquad \bm n_n=\frac{c\bm k_n}{z}.
 \label{S:orbit-operator}
\end{equation}
The off-diagonal blocks with $m\ne n$ encode nonlocal coupling from the density gradient and finite gyro-orbits.  The dimensionless momentum $q_\parallel=p_\parallel/(m_sc)$ is integrated along $q_\parallel=\sqrt{2\Theta_{sa}}(y-\ii d)$, with $d>0$.  Moving this contour within the same retarded domain changes the pole only at quadrature-error level.

Three independent tests constrain the implementation.  In the homogeneous
limit, the orbit kernel agrees with the separately evaluated Maxwell--J\"uttner
tensor to a maximum relative matrix error of $9.4\times10^{-12}$.  The
discrete charge and current responses satisfy the Ward relation
$z\rho=\bm k_{\rm out}\cdot\bm J$ with relative residual
$6.2\times10^{-10}$.  Finally, the orbit operator differs from the frozen
local operator by $2.7\times10^{-5}$ in the same Fourier box, consistent with
the small orbit-width-to-interface-width ratio.

At $u_y=0$ and $u_z=1$, the finest discretization used here gives
\begin{equation}
 w_{\rm orbit}\simeq0.6126682-2.9789\times10^{-3}\ii .
 \label{S:orbit-pole}
\end{equation}
The Keldysh denominator defined in Eq.~(\ref{S:keldysh-denominator}) is nonzero
and stable under derivative-step refinement, so Eq.~(\ref{S:orbit-pole}) is a
simple rank-one response pole.  Its central-interface weight is $0.9371$, its
periodic-seam weight is $5.1\times10^{-7}$, and its rms width is
$\sigma_X=0.1362$.  Relative to the principal-symbol pole in
Eq.~(\ref{S:rel-edge}), the correction is
\begin{equation}
 \Delta w=w_{\rm orbit}-w_{\rm edge}^{\rm rel}
 =(4.15+3.67\ii)\times10^{-6}.
 \label{S:orbit-shift}
\end{equation}
A five-node homotopy between the local Fourier-box operator and
$\mathsf T^+_{\rm orbit}$ follows the same simple pole at every node.  No
state switch or loss of interface localization occurs.

Because a complex nonlinear eigenvector has an arbitrary global phase, we
report the phase-invariant component amplitudes
$|E_\mu(X)|/\max_X|\bm E(X)|$.  The integrated electric-field fractions
\begin{equation}
 P_\mu=\frac{\sum_i|E_{i\mu}|^2}
 {\sum_{i,\nu}|E_{i\nu}|^2},\qquad \mu\in\{x,y,z\},
 \label{S:orbit-polarization}
\end{equation}
are
$(P_x,P_y,P_z)=(0.1988,0.1937,0.6074)$.  The total-field amplitude peaks at $X=0.225$, adjacent to the physical interface at $X=0.2211$ [Fig.~\ref{fig:S-orbit-connection}(d)].  Its polarization is predominantly field aligned, with $E_z$ carrying $60.7\%$ of the integrated electric-field intensity.

An orbit-resolved scan over $-0.009\le u_y\le0.009$ gives
\begin{equation}
 \frac{dw_{\rm orbit}}{du_y}=0.28298-0.01569\ii .
\end{equation}
The branch crosses $\operatorname{Re}w=\operatorname{Re}w_{\rm EP}^{\rm rel}$ once upward and never downward, so its finite-window oriented count is $+1$.  The finite-window equality $\nu_{\mathcal W}=C^-_{\rm F}$ therefore survives the leading nonlocal orbit correction.  The minimum central-interface weight is $0.936$, and the largest periodic-seam weight is $8.4\times10^{-5}$.  This calculation tests the boundary response beyond the frozen principal symbol; it does not define a Chern class for the full nontranslation-invariant operator.

\subsection{Wide-window outgoing-pole diagnostic}

As a wider-window diagnostic, we compare a local outgoing Jost pole with the exact bulk spectral sets.  The propagation model uses the first kinetic $u_x$ jet and the exact quadratic Maxwell block.  Every graph-Evans zero is checked through the minimum singular value of the matching matrix.  The comparison roots are solved from the exact retarded dispersion for real $u_x\in[-1.5,1.5]$.

\begin{figure}[H]
 \centering
 \includegraphics[width=\textwidth]{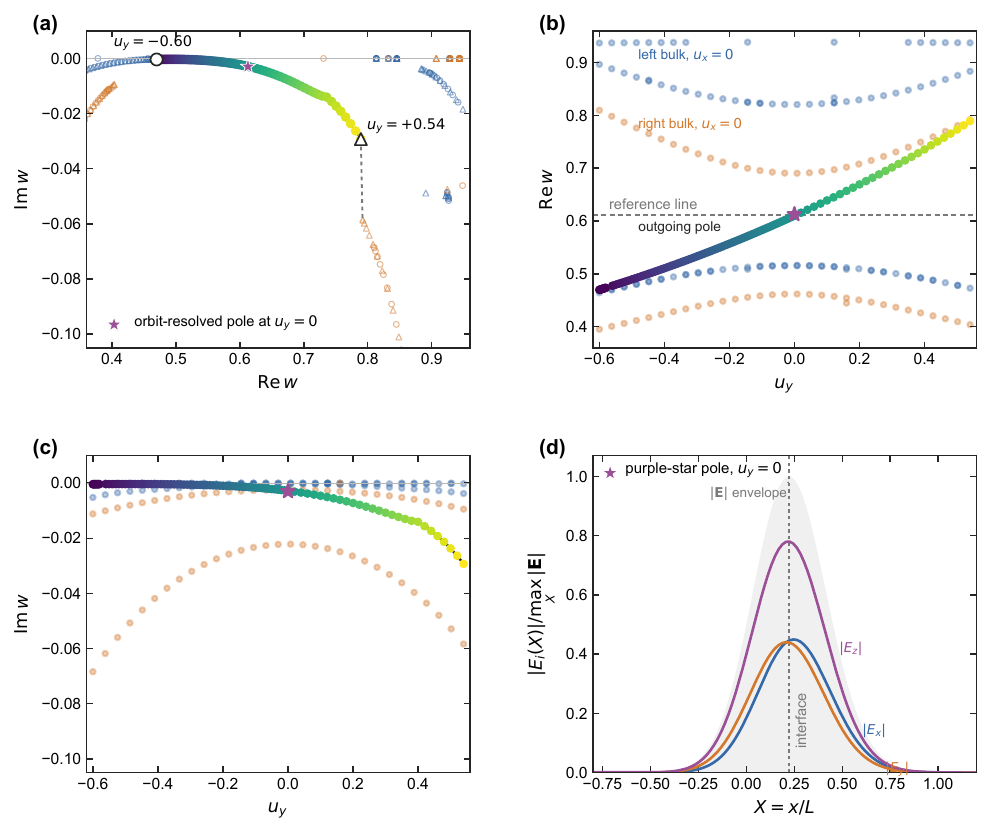}
 \caption{\label{fig:S-orbit-connection}Wider-window crossing signature and
 orbit-resolved interface mode.  (a) Complex-frequency trajectory of the
 singular-value-validated outgoing pole as $u_y$ is varied.  Open symbols are
 exact full-$u_x$ retarded bulk roots at the two ends of the sampled pole
 interval; color distinguishes the two asymptotic density plateaus and symbol
 shape distinguishes the endpoint values of $u_y$.  Dashed segments join each
 pole endpoint to its nearest sampled exact-bulk root.  The purple star is the
 independently computed orbit-resolved pole at $u_y=0$.  (b,c) Real and
 imaginary parts of the outgoing pole, with $u_x=0$ cuts of the exact left and
 right bulk spectra.  The pole crosses the reference line once.  (d)
 Phase-invariant amplitudes $|E_x|$, $|E_y|$, and $|E_z|$ of the purple-star
 orbit-resolved mode, normalized by $\max_X|\bm E|$.  The gray envelope is
 $|\bm E|/\max_X|\bm E|$, and the dashed line marks the physical interface.
 The profile is localized and $E_z$ carries $60.7\%$ of the integrated
 electric-field intensity.}
\end{figure}

The outgoing pole remains continuous over $-0.60\le u_y\le0.54$ and crosses the reference line once [Fig.~\ref{fig:S-orbit-connection}(a--c)].  Its endpoint distances to the nearest exact full-$u_x$ bulk roots are $5.5\times10^{-3}$ and $2.9\times10^{-2}$, and the spatial decay clearances remain finite.  At $u_y=0$, the local result $w_{\rm Jost}=0.61008-2.77\times10^{-3}\ii$ differs from the orbit-resolved pole by $2.6\times10^{-3}$.  The calculation therefore provides a wider-window crossing diagnostic, but does not demonstrate attachment to both exact bulk spectral sets.

\section{Scope of the semiclassical claim}

Equation~(\ref{S:eq}) is an exact nonuniform nonrelativistic equilibrium, and Eq.~(\ref{S:rel-equilibrium}) is its self-consistent relativistic counterpart.  The full collisionless Vlasov--Maxwell evolution remains Hamiltonian and conservative~\cite{Morrison1980,Ramos2019}.  The complex roots belong to the retarded response obtained after eliminating velocity or momentum space.

The reported Chern class belongs to the retarded Wigner principal-symbol root bundle.  The nonrelativistic cold--hot continuation and relativistic homotopy have the same scope.  Equation~(\ref{S:orbit-operator}) tests the leading nonlocal boundary correction through exact gyro-orbit sampling.  It yields a simple localized pole, a smooth local-to-orbit continuation, and one oriented crossing in the stated window.  No Chern class is assigned to the full nontranslation-invariant operator or to the conservative Vlasov generator.

At the kinetic endpoint, the computed relation is $\nu_{\mathcal W}=C^-_{\rm F}=+1$ within the stated line-gap window.  Faure's theorem establishes the corresponding cold Hermitian relation.  The bulk homotopy preserves the charged root bundle, while the $u_y=0$ pole homotopy connects the cold mode to the hot response pole.  Together with the hot scan, these results motivate an index principle for nonlinear retarded pencils.  Establishing such a theorem would require control over the full temperature--momentum homotopy and possible continuum or purely kinetic branches.  The exact-bulk comparison in Fig.~\ref{fig:S-orbit-connection} remains an ancillary diagnostic rather than a global bulk-attachment result.

\section{Data and code availability}

The reproducibility repository contains the Python programs, JSON source data, figure scripts, and dependency specification.  Nonrelativistic code is under \path{code/}, principal-symbol data under \path{data/}, the Maxwell--J\"uttner extension under \path{relativistic_extension/}, and the orbit kernel and outgoing calculation under \path{orbit_resolved_relativistic/}.  All commands are relative to the repository root.

The archive covers the radius scaling, Chern meshes, bulk and boundary continuations, Keldysh and contour tests, and relativistic and orbit-resolved checks.  A clean-clone test regenerates both main figures and recomputes a reduced $N_X=16$ pole without absolute paths.  Numerical cutoffs, operator ordering, species content, causal contours, and derivative steps are exposed as inputs.  Version 1.0.0 is archived in Zenodo at \href{https://doi.org/10.5281/zenodo.22020911}{doi:10.5281/zenodo.22020911}; code is licensed under BSD-3-Clause and machine-readable data under CC BY 4.0.

\bibliography{references}